\documentclass[twocolumn]{aastex631}

\usepackage{natbib}
\usepackage{gensymb}
\usepackage{multirow}

\newcommand{\radm}{{\rm rad\ m^{-2}}}

\shorttitle{RM jitter}
\shortauthors{J. M. Stil}

\begin{document}

\title{RM jitter:\\ Wavelength-dependent Scatter in Rotation Measure Related to Faraday Complexity}

\correspondingauthor{Jeroen Stil}
\email{jstil@ucalgary.ca}

\author{Jeroen M. Stil}
\affiliation{Department of Physics and Astronomy, The University of Calgary \\
2500 University Drive NW \\
Calgary AB T2N 1N4, Canada}

\begin{abstract}
The relation between Faraday Rotation Measure ($RM$) and differential Faraday rotation by unresolved structure of a turbulent plasma is investigated for extragalactic radio sources. The $RM$ scatter of a sample of sources affected by turbulent Faraday screens with identical power spectra of Faraday depth structure, is referred to as $RM$ jitter. For fixed amplitude and slope of the power spectrum, the range of possible $RM$s depends on the wavelength coverage of the survey. $RM$ jitter is independent of Faraday depth resolution as it results from the true Faraday depth dispersion and effects of wavelength-dependent depolarization. $RM$ jitter for a flux density limited sample is sensitive to the power law index $\gamma$ of the power spectrum of Faraday depth structure. Assuming depolarization by a turbulent Faraday screen for all sources, a simulated flux-density-limited sample can reproduce the high $RM$ scatter found by \citet{vanderwoude2024} for sources that are less than $3\%$ polarized. $RM$ jitter of sources that are more than 3\% polarized, is found to be smaller than the observed scatter, indicating that plasma other than the near-source environment dominates the $RM$ scatter for the more polarized sources. The significance of $RM$ jitter for applications of the $RM$ grid is discussed.
\end{abstract}


\section{Introduction} \label{intro-sec}

Faraday rotation is the only observational tool to probe galactic scale magnetic fields in astrophysical plasmas that are not illuminated by synchrotron emission of relativistic electrons, and it is one of the most powerful tools to map magnetic fields in plasmas where the magnetic field is illuminated by relativistic electrons. One of the most versatile data sets for investigation of cosmic magnetism at low and at high redshift is a Rotation Measure grid of extragalactic sources \citep{beck2004}. As Square Kilometre Array (SKA)\footnote{https://skao.int} pathfinders have begun operations, and construction of the SKA is underway, connecting Faraday dispersion arising from turbulence in a plasma near the source with observable quantities such as fractional polarization and $RM$ scatter is an important step in developing its key science goal Origin and Evolution of Cosmic Magnetic Fields \citep{johnstonhollitt2015,heald2020}.

A linearly polarized radio wave that travels through a magnetized plasma experiences rotation of the plane of polarization, traced by the polarization angle $\theta$, according to
\begin{equation}
\theta - \theta_0 = \phi \lambda^2,
\label{FR_simple-eq}
\end{equation}
where $\theta_0$ is the polarization angle at the source, $\lambda$ is the wavelength of the wave at the time it travels through the plasma, and $\phi$ is the Faraday depth defined as
\begin{equation}
\phi = {e^3 \over 2 \pi m_e^2 c^4} \int n_e B_\| dl,
\label{phi-eq}
\end{equation}
with $e$ the elementary charge, $c$ the speed of light, $m_e$ the mass of an electron, $n_e$ the density of free electrons, and $B_\|$ the component of the magnetic field along the line of sight. The total observable linearly polarized signal, expressed in terms of the complex linear polarization $\mathcal{P} = q + i u$, with $q=Q/I$ and $u=U/I$ the Stokes parameters for linear polarization divided by the total flux density, is \citep{brentjens2005}
\begin{equation}
\mathcal{P} (\lambda^2)= \int_{-\infty}^{\infty} \mathcal{F}(\phi) e^{2i \phi \lambda^2} d \phi,
\label{RMsynth_inv-eq}
\end{equation}
where the complex Faraday dispersion function $\mathcal{F}(\phi)$ represents the distribution of linearly polarized emission in $\phi$, integrated over the volume of space defined by the beam and the line of sight. 

In $RM$ synthesis \citep{brentjens2005}, the sampling of $\mathcal{P}(\lambda^2)$ by a survey is expressed in terms of a weight function $W(\lambda^2)$, which is zero where no measurements exist, and normalized to have its integral equal to 1. The inverse Fourier transform of the available data leads to the observed Faraday dispersion function,
\begin{equation}
\tilde{\mathcal{F}}(\phi) = \int_{-\infty}^{\infty} \mathcal{P}(\xi) W(\xi) e^{-2i \phi \xi} d \xi,
\label{RMsynth-eq}
\end{equation}
with $\xi = \lambda^2$ when $\xi > 0$ and $W(\xi) = 0$ when $\xi \le 0$ \citep{brentjens2005}. The inverse Fourier transform of $W$ is called the Rotation Measure Transfer Function ($RMTF$). The Rotation Measure $RM$ is defined as the peak of $|\tilde{\mathcal{F}}|$, or its reconstructed equivalent after $RM$ Clean \citep{heald2009}. The measurement error of $RM$ for a source with polarized flux density that is a multiple $M$ of the noise $\sigma$ in polarization is
\begin{equation}
\Delta_{M\sigma} = {\Delta \phi_{\rm FWHM} \over 2M}
\label{RM_error-eq}
\end{equation}
\citep{brentjens2005}, where $\Delta \phi_{\rm FWHM}$ is the full width at half of the peak of $|RMTF|$. Surveys with a large bandwidth have in recent years reduced $\Delta_{M\sigma}$ by an order of magnitude compared with older, narrow band surveys. This has greatly increased the scientific potential of observed scatter in $RM$, and the significance of understanding it in more detail.  

We also define the quantity
\begin{equation}
\mathcal{R} = {d \theta \over d \lambda^2}.
\label{RM-eq}
\end{equation}
Although $\mathcal{R}$ is often identified with $RM$, the definition of $\mathcal{R}$ is monochromatic while the definition of $RM$ involves, potentially, averaging over a wide wavelength range (Equation~\ref{RMsynth-eq}). This subtle difference is important here. 

$RM$ synthesis is a non-parametric algorithm to create a polarization catalog from survey data. As such, it is the main analysis method in this paper. $QU$ fitting \citep{law2011} is a model-dependent alternative where $\mathcal{P}(\lambda^2)$ is fitted with a physically motivated model. A popular model from \citet{burn1966} and \citet{sokoloff1998},
\begin{equation}
\mathcal{P}(\lambda^2) = |\mathcal{P}_0|  \exp[-2 \sigma_\phi^2 \lambda^4] \exp[2i (\theta_0 + \phi_0 \lambda^2)],
\label{depol_model-eq}
\end{equation}
treats the Faraday rotation parameter $\phi_0$ as independent from the Faraday depth dispersion, $\sigma_\phi$, which causes only depolarization. This model applies to depolarization by a foreground screen with structure much smaller than the scale of the beam, or the source, whichever is smaller. The net Faraday rotation could be from the screen itself, or another screen, as in the case of an extragalactic source behind the Milky Way \citep{woltjer1962}.

Another popular model for $QU$ fitting considers synchrotron emission within a plasma with uniform magnetic field and electron density,
\citep{sokoloff1998,osullivan2017}
\begin{equation}
\mathcal{P} = |\mathcal{P}_0|  {\sin \Phi \lambda^2 \over \Phi \lambda^2 } \exp[2i (\theta_0 + {1 \over 2}\Phi \lambda^2)],
\label{burnslab-eq}
\end{equation}
where the parameter $\Phi$ is the total Faraday depth of the source, and $RM = \Phi/2$. Here we see a connection between $RM$ and depolarization, as both result from the same integration of Faraday rotated emission along the line of sight. In both models, Equation~\ref{depol_model-eq} and Equation~\ref{burnslab-eq}, $\mathcal{R}$ is a constant \citep{sokoloff1998}, and $RM = \mathcal{R}$, except when complete depolarization occurs in Equation~\ref{burnslab-eq}. The interpretation of $RM$ in terms of the total Faraday depth of the plasma is different for these models. Quantifying Faraday depth dispersion from survey data remains an area of active research \citep[e.g.][]{sun2015,anderson2015,carcamo2023}.

\citet{tribble1991} investigated depolarization by a turbulent screen that has structure on a range of scales, and derived an expression for depolarization that varies between the exponential form in Equation~\ref{depol_model-eq} and a power law dependence on wavelength. As radio lobes expand into the circumgalactic medium of the host galaxy, it is not unreasonable to expect that the energy injection scale for turbulence is roughly the scale of the lobes. \citet{tribble1991} was mainly concerned with median depolarization properties, but also noted that the largest structures within the beam result in a certain amount of randomness. \citet{shanahan2023} showed the range of depolarization curves that can arise from different realizations of the same model.

This paper presents an analysis of simulations similar to the numerical experiments of \citet{tribble1991}, applied to modern broad-band surveys analyzed with $RM$ Synthesis. Contrary to the approach of \citet{tribble1991}, the stochastic nature of these models is paramount here\footnote{Tribble did consider $RM$ scatter in relation with the angular size of the source, compared with the beam.}, in particular the breakdown of linearity of $\theta(\lambda^2)$ discussed in Figure 6 of \citet{tribble1991} when $\sigma_\phi \lambda^2 >1$ in the quasi-monochromatic treatment of \citet{tribble1991}. The focus of this paper is the $RM$ scatter in a sample of extragalactic radio sources by the plasma that also causes depolarization by differential Faraday rotation across the source.

\section{$RM$ jitter}
\label{RMjitter-sec}

\subsection{Simulations}

\begin{figure*}
\centerline{\resizebox{12cm}{!}{\includegraphics[angle=0,trim={0.1cm 1.0cm 0.1cm 3.2cm},clip]{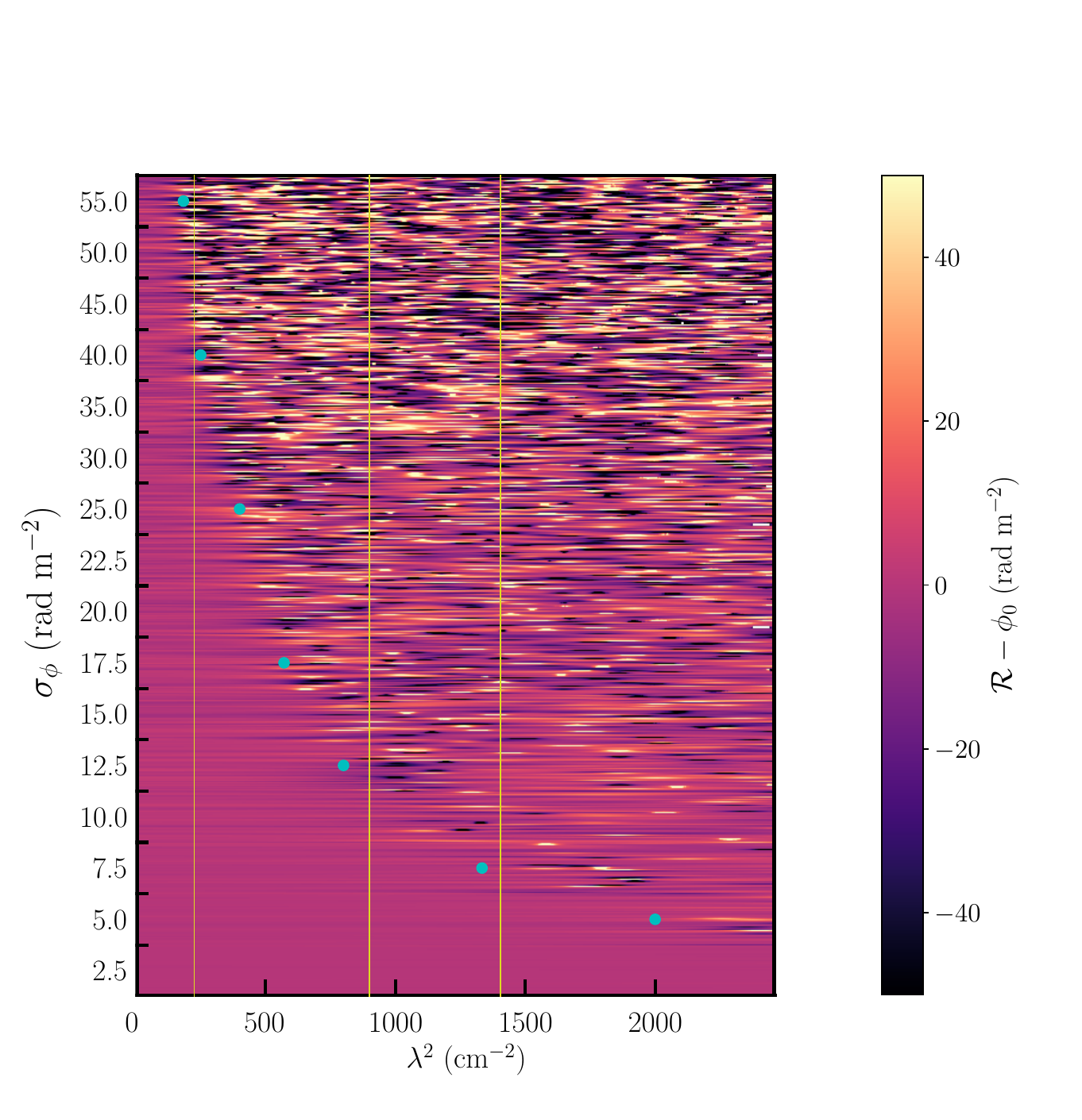}}}
\caption{ $\mathcal{R}$ defined in Equation~\ref{RM-eq} as a function of $\lambda^2$ and $\sigma_\phi$. The constant $\phi_0$ was subtracted for clarity. For the $\sigma_\phi$ axis, 50 simulated sources are displayed from bottom to top for each of $\sigma_\phi = $ 2.5, 5.0, 7.5, 10.0, 12.5, 15.0, 17.5, 20.0, 22.5, 25.0, 30.0, 35.0, 40.0, 45.0, 50.0, and 55.0 $\radm$. 
Cyan dots outline the boundary where $\sigma_\phi \lambda^2 = 1$. The vertical lines mark, from left to right, frequencies 2.0 GHz, 1.0 GHz, and 0.8 GHz. There is no noise in these simulations. Approximately 2\% of values are outside the display range, equally divided between positive and negative sides.
\label{gradient-fig}}
\end{figure*}

A turbulent plasma may have structure on a range of scales from much smaller to much larger than the beam or source, whichever is smaller.  Structures on much larger scales contribute mainly to the $RM$. Structures on much smaller scales contribute mainly to Faraday depth dispersion, $\sigma_\phi$. Structures on intermediate scales do not average out completely when integrating over the beam. They link $RM$ to the Faraday depth dispersion. 

Simulations of radio sources with unresolved Faraday depth structure are generated numerically by assuming a power spectrum of Faraday depth structure, $\mathcal{A}(k)$, with $k$ the wave number, of the form
\begin{equation}
\mathcal{A}(k) = A k^\gamma.
\label{powerspec-eq}
\end{equation}
The exponent $\gamma$ was taken to be $-2.0$, $-2.5$, and $-3.0$, which is in the range suggested by observations of well-resolved radio galaxies \citep{laing2008}. A 2-dimensional hermitian complex function with amplitudes defined by Equation~\ref{powerspec-eq} is then generated by including random phases. The desired random field is found by Fourier transformation and multiplication with a positive constant that rescales the constant $A$ in Equation~\ref{powerspec-eq}. Every simulation was scaled to make the standard deviation of $\phi$ across the source, which is $\sigma_\phi$, equal to a prescribed value. After scaling the random field, the constant $\phi_0 = 300\ \radm$ is added to include a non-zero mean $RM$. The only purpose for this constant is visualization of the simulations. It has no effect on the analysis. 

\begin{figure*}
\resizebox{9cm}{!}{\includegraphics[angle=0,trim={0.1cm 0.0cm 0.1cm 1.5cm},clip]{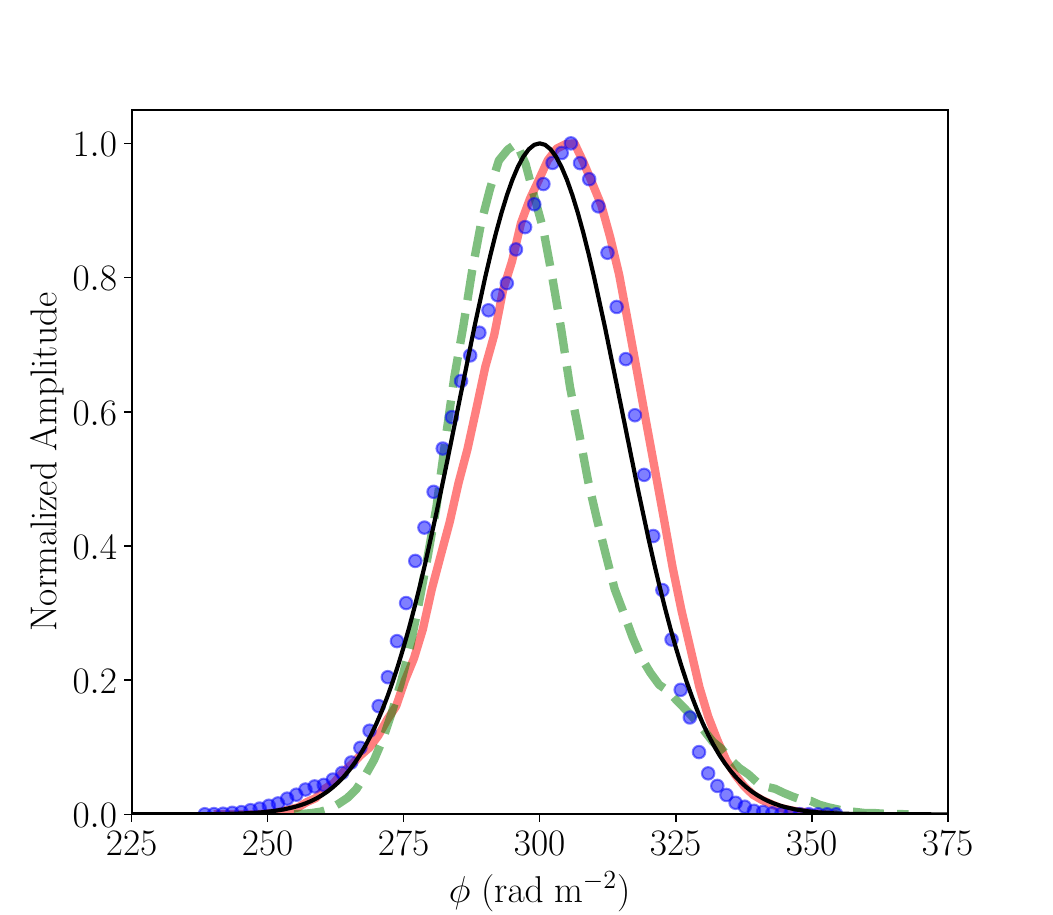}}
\resizebox{9cm}{!}{\includegraphics[angle=0,trim={0.1cm 0.0cm 0.1cm 1.5cm},clip]{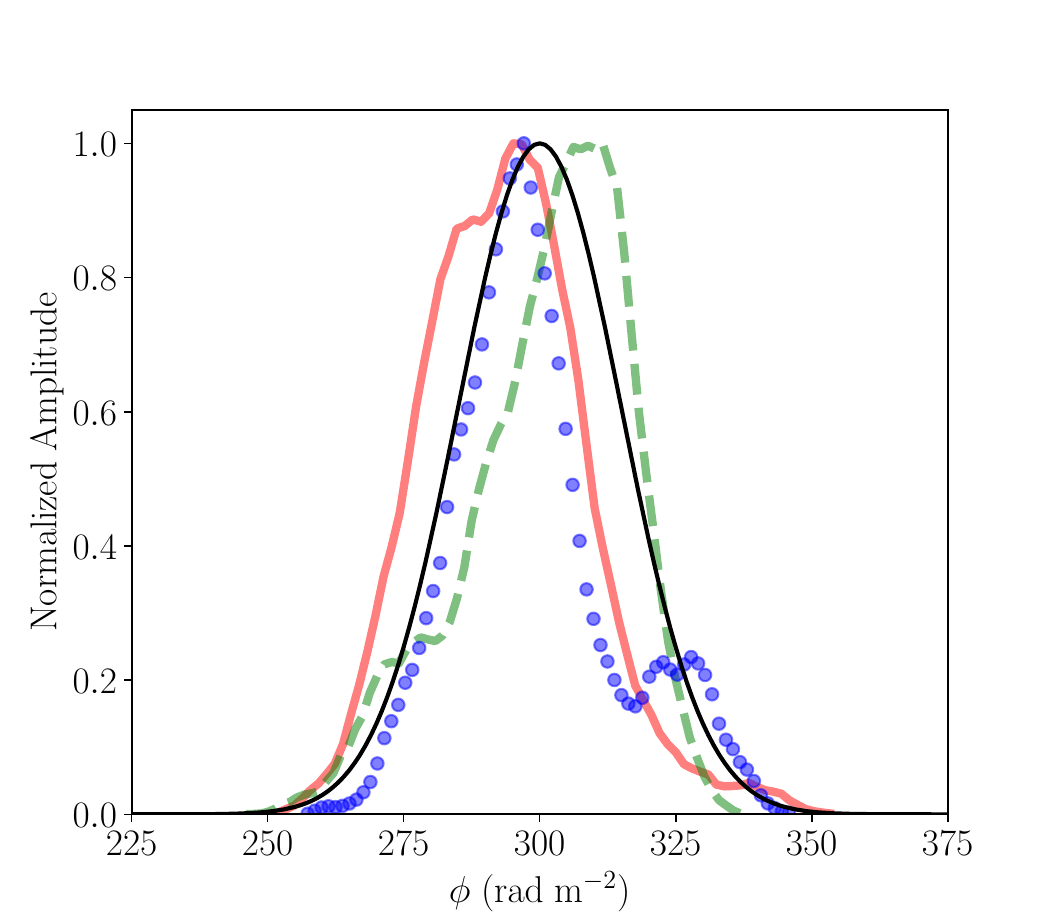}}
\caption{ Representative distributions of Faraday depth in simulated sources. The left panel shows three simulations with $\gamma=-2.5$, and, for reference, a Gaussian distribution with mean $300.00\ \radm$ and standard deviation $\sigma_\phi = 15.00\ \radm$. The red continuous distribution has mean $302.39\ \radm$ (also shown in Figure~\ref{RMsynth-fig}), the green dashed distribution has mean $298.59\ \radm$, and the blue dotted distribution has mean $299.30\ \radm$. All distributions in both panels have standard deviation $\sigma_\phi = 15.00\ \radm$. The right panel shows three simulations with $\gamma=-3.0$, and the same Gaussian distribution as shown in the left panel (black curve). The red continuous distribution has mean $293.72\ \radm$, the green dashed distribution has mean $301.10\ \radm$, and the blue dotted distribution has mean $299.08\ \radm$.
\label{phi_dist-fig}}
\end{figure*}

Figure~\ref{phi_dist-fig} shows examples of Faraday depth distributions in simulations with $\gamma=-2.5$ and $\gamma=-3.0$. The continuous red distribution in the left panel is also used in Figure~\ref{RMsynth-fig}. The steeper the slope of the power spectrum, the stronger the deviations from a Gaussian distribution. When the power spectrum is steep, large scale structure appears more prominently as broad "Faraday components". The colored distributions in each panel in Figure~\ref{RMsynth-fig} represent turbulent screens with \textit{the same} power spectrum. The differences occur because the large-scale structure defines a small number of "cells" across the source with somewhat coherent Faraday rotation. While the cumulative effect of small-scale structure may appear Gaussian because of the central limit theorem, this cannot apply, by definition, to the largest scales, which are more prominent if the power spectrum is steep and extends to these larger scales.

The sampling in the Fourier domain includes scales (wave numbers) that vary from the size of the generated image (essentially the diameter of the circular source) down to the pixel scale. The simulations therefore include the scales that lead to differential Faraday rotation across the source, causing wavelength-dependent depolarization. It is not within the scope of this paper to explore in detail a cut-off of the power spectrum at larger scales. Including scales larger than the source would add $RM$ scatter unrelated to differential Faraday rotation across the source, which, for the purpose of this paper, is part of the foreground $RM$. A small-$k$ cut-off of the power spectrum on angular scales smaller than the source would create models that resemble Equation~\ref{depol_model-eq} more closely.

Faraday rotation by unresolved turbulent structure can then be calculated by associating each pixel in the image with a Faraday depth $\phi$ and evaluating a sum over all pixels that approximates the integral over the source. For simplicity, the source is a circular area with radius 400 pixels, for approximately a half million individual pixels in the source (Figure~\ref{RMsynth-fig}). Stokes $Q$ and $U$ were evaluated according to Equation~\ref{FR_simple-eq}, with $\theta_0=0$, for every pixel and integrated over the source to obtain the complex polarization $\mathcal{P}(\lambda^2)$. 

Real sources have a distribution of $\theta_0$ that leads to wavelength-independent depolarization when the source is unresolved. This defines the location of the source in the $QU$ plane in the short-wavelength limit. A uniform distribution of $\theta_0$ leads to an unpolarized source at all wavelengths. If one considers some distribution of $\theta_0$ that leads to structure in $Q$ and $U$ at the shortest wavelengths, that structure will remain unaltered, with near constant integrated polarization $\mathcal{P}$, until significant differential Faraday rotation occurs. In models with $\theta_0=0$, differential Faraday rotation creates a polarization angle distribution at finite but short wavelength. Imposing an explicit $\theta_0$ distribution only has a moderate effect by redefining the initial conditions of the model. It does require additional parameters that, while interesting, do not relate directly to the essence of this paper.

The effect of angular resolution on the models is minimal as most sources fit well within the beam of current wide-area surveys. The simulations include a Gaussian synthesized beam that is 10 times the diameter of the source, corresponding to a survey with a synthesized beam of $20\arcsec$ and a source diameter of $2\arcsec$. The synthetic $\mathcal{P}(\lambda^2)$ were evaluated in the frequency range $600$ MHz to $8$ GHz at 2 MHz intervals. The frequency channels are assumed so narrow that differential Faraday rotation within a channel is negligible. 

Figure~\ref{gradient-fig} shows $\mathcal{R}$ as a function of $\lambda^2$ and $\sigma_\phi$ for simulations with $\gamma=-2.5$. For each value of $\sigma_\phi$, a subset of 50 independent simulated sources is shown. Constant $\mathcal{R}$ indicates a linear relation $\theta(\lambda^2)$. At longer wavelengths and higher $\sigma_\phi$, $\mathcal{R}$ displays erratic structure with amplitude of tens, sometimes hundreds $\radm$, much more than the Faraday depth dispersion across the source, $\sigma_\phi$. The boundary $\sigma_\phi \lambda^2$ is marked by cyan dots. This sharp boundary occurs naturally in the simulations. \citet{tribble1991} (his Figure 6) discussed it as the transition between linear $\theta(\lambda^2)$ and non-linear $\theta(\lambda^2)$. 

At shorter wavelengths, $\lambda^2 < 1/\sigma_\phi$, there are source-to-source differences in $\mathcal{R}$, but they are small and nearly independent of wavelength.  The source-to-source variation of $\mathcal{R}$ increases with $\sigma_\phi$ in the regime $\lambda^2 < 1/\sigma_\phi$, as well in the regime $\lambda^2 > 1/\sigma_\phi$, where the dependence of $\mathcal{R}$ on $\sigma_\phi$ is strongest. 

Figure~\ref{gradient-fig} clearly demonstrates the need to distinguish between $\mathcal{R}$ and $RM$, because the latter may average significant structure in $\theta(\lambda^2)$ during $RM$ synthesis or $QU$ fitting. For narrow band surveys, $RM$ approaches $\mathcal{R}$. Figure~\ref{gradient-fig} therefore shows how $RM$ scatter depends on $\sigma_\phi$ and wavelength for samples of unresolved sources in narrow band surveys. It will now be shown that these trends also exist for sources observed in wide-band surveys. 

\subsection{Application to wide-band surveys}
 
A selected set of surveys is defined in Table~\ref{surveys-tab}. Parameters for the Polarization Sky Survey of the Universe's Magnetism (POSSUM) were taken from \citet{vanderwoude2024}. The POSSUM Low survey represents expectations for most of the POSSUM survey, while the POSSUM Low $+$ Mid survey may represent a smaller subset \citep{vanderwoude2024}. Parameters for The HI/OH/Recombination line (THOR) survey were adapted from \citet{shanahan2023}, and parameters for a narrow-band survey labeled NVSS/CGPS were taken from \citet{landecker2000} to jointly represent major narrow band surveys outside and within the Galactic plane. Sparsely sampled surveys of narrow-band observations over a large frequency range \citep[e.g.][]{simard-normandin1981,klein2003,farnes2014a} are not considered here, because these surveys must rely on a linear relation $\theta(\lambda^2)$ to resolve the $n \pi$ ambiguity of the polarization angle \citep[e.g.][]{simard-normandin1981}. The $RM$ scatter in the resulting catalog will depend on details of this procedure and possible rejection criteria if no acceptable solution can be found for a source. Three mock surveys were included for the purpose of discussion: B001, Wide, and B002. The frequency sampling of the models is 2 MHz channels for every survey.  

Table~\ref{surveys-tab} also lists the resolution in Faraday depth, $\Delta \phi_{\rm FWHM}$, from a fit to $|RMTF|$ by $RM$-tools\footnote{https://github.com/CIRADA-tools/RM} \citep{purcell2020}, and the largest continuous Faraday depth range detectable, using the parameter introduced by \citet{rudnick2023},
\begin{equation}
W_{\rm max} = 0.67(\lambda_{\rm max}^{-2}+\lambda_{\rm min}^{-2}),
\label{depol-eq}
\end{equation} 
with $\lambda_{\rm max}$ the maximum wavelength and $\lambda_{\rm min}$ the minimum wavelength. $W_{\rm max}$ represents a continuous range of Faraday depth that leads to a depolarization factor $1/2$, whereas the depolarization parameter introduced by \citet{brentjens2005} refers to complete depolarization. Table~\ref{surveys-tab} further lists the mean $\lambda^2$ calculated by $RM$-tools, $\lambda_0^2$, and the expected measurement error of $RM$ for a source detected at $10\sigma$ in polarization, $\Delta_{10\sigma}$ calculated as $\Delta \phi_{\rm FWHM}/20$ (Equation~\ref{RM_error-eq} for $M = 10$). The B001 survey was designed to have the same $W_{\rm max}$ as the POSSUM Low survey, but $\sim 3$ times poorer Faraday depth resolution.

\begin{center}
\begin{table*}[!ht]
\caption{Survey parameters}
\begin{tabular}{cccccc}
\hline\hline
Name & Frequency range &  $\Delta \phi_{\rm FWHM}$ & $W_{\rm max}$  & $\lambda_0^2$  & $\Delta_{10\sigma}$ \\
           &   (MHz)     &  $\radm$ & $\radm$  &   $\rm cm^2$  & $\radm$ \\
\hline
POSSUM Low       &                            800 - 1087         &  59.8            &    13.6      & 1032.2  & 3.0  \\
B001                    &        900 - 1006 &    171.6       &   13.6  &  992.8  & 7.8 \\
POSSUM Low $+$ Mid        & 800 - 1087, 1316 -  1439     & 39.5 &  20.2 & 865.4    & 2.0 \\ 
NVSS/CGPS         &                  1405 - 1435        & 2161              &  30.1    & 445.8   &  108 \\
THOR        & 1000 - 1128, 1256 - 2000    & 56.7       &    37.27       &  421.4   &  2.8 \\
Wide           &  600 - 8000                         &  15.3      &    479.8       &  187.6 &  0.70 \\
B002           &     3000 - 8000                    &  442.8    &    544.2       &    37.47    &  20.2  \\
\hline
\end{tabular}
\label{surveys-tab}
\end{table*}
\end{center}

\begin{figure*}
\resizebox{10cm}{!}{\includegraphics[angle=0,trim={1.3cm 0 2.0cm 2.8cm},clip]{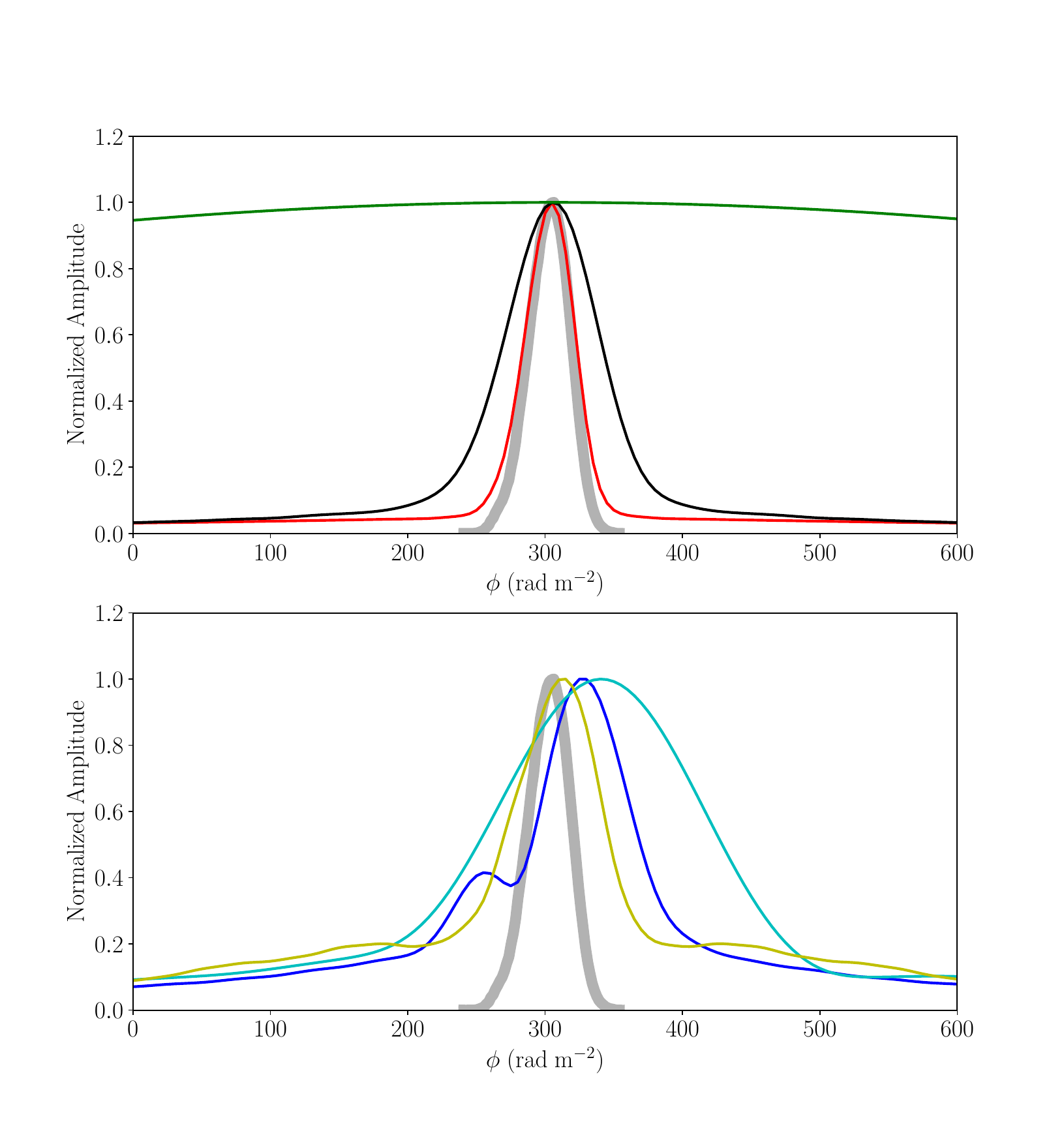}}
\resizebox{6cm}{!}{\includegraphics[angle=0,trim={0.1cm 0 1.5cm 1.8cm},clip]{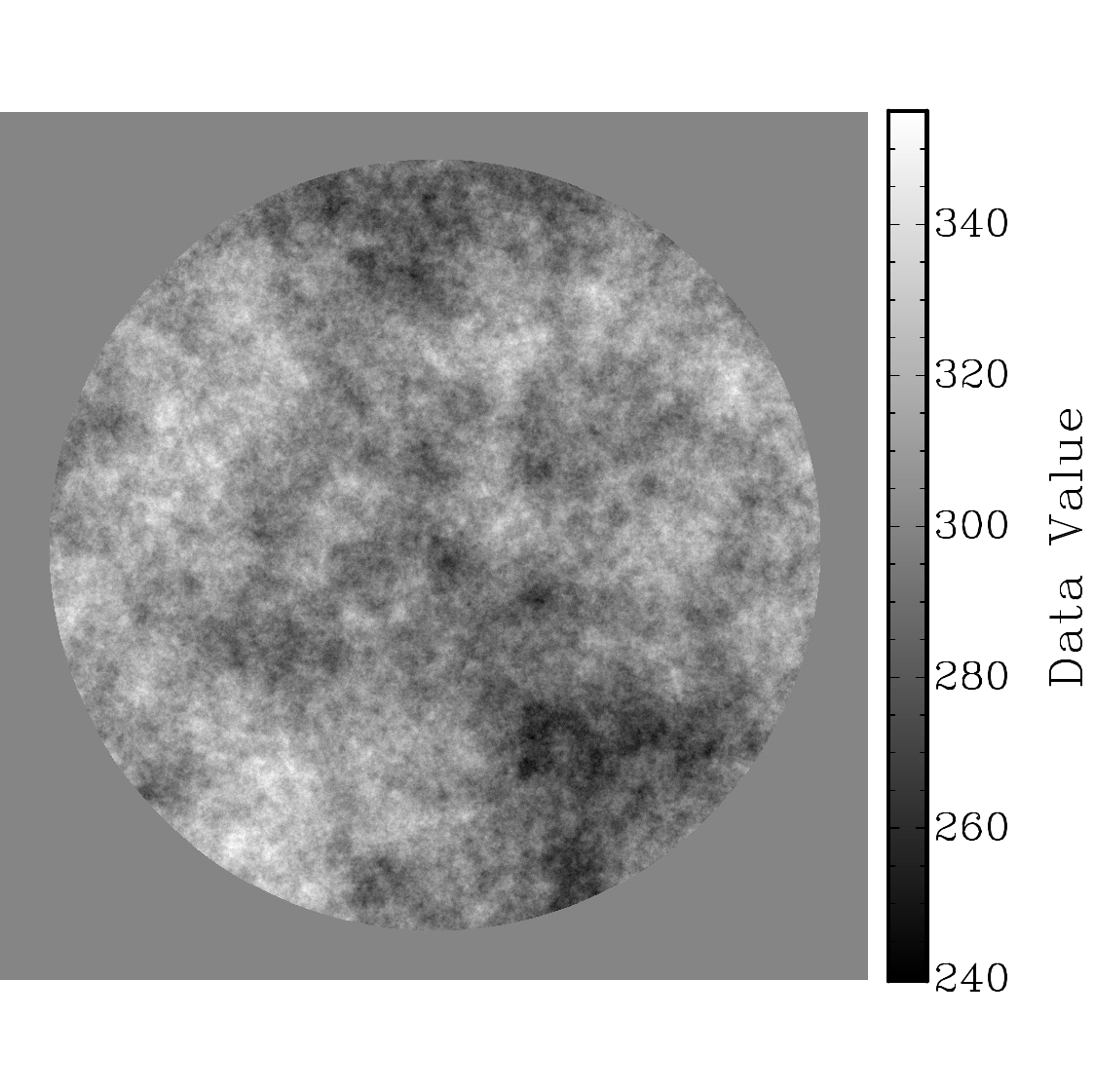}}
\caption{ Deconvolved Faraday dispersion function for 6 surveys of a single simulated source with Faraday depth distribution as shown on the right, with $\sigma_\phi=15\ \radm$ and $\gamma=-2.5$. The histogram of the model Faraday depths is drawn as a thick gray curve, with full range $240.0\ \radm$ to $354.8\ \radm$. Top panel: THOR survey (black, $RM=306.48\ \radm$), NVSS/CGPS survey (green, $RM=306.24\ \radm$), and full simulated wavelength range of 600 MHz to 8 GHz (red, $RM = 303.85\ \radm$). The mean Faraday depth for the source is $302.39\ \radm$. Bottom panel: POSSUM Low (blue, $RM=326.68\ \radm$), POSSUM Low $+$ Mid (yellow, $RM=315.96\ \radm$), B001 (cyan, $RM=340.77\ \radm$). This simulation was selected as a more extreme positive deviation for the POSSUM survey for demonstration purposes. 
\label{RMsynth-fig}}
\end{figure*}

\begin{figure*}
\centerline{\resizebox{12.0cm}{!}{\includegraphics[angle=0,trim={0.0cm 1.0cm 0.0cm 2.5cm}]{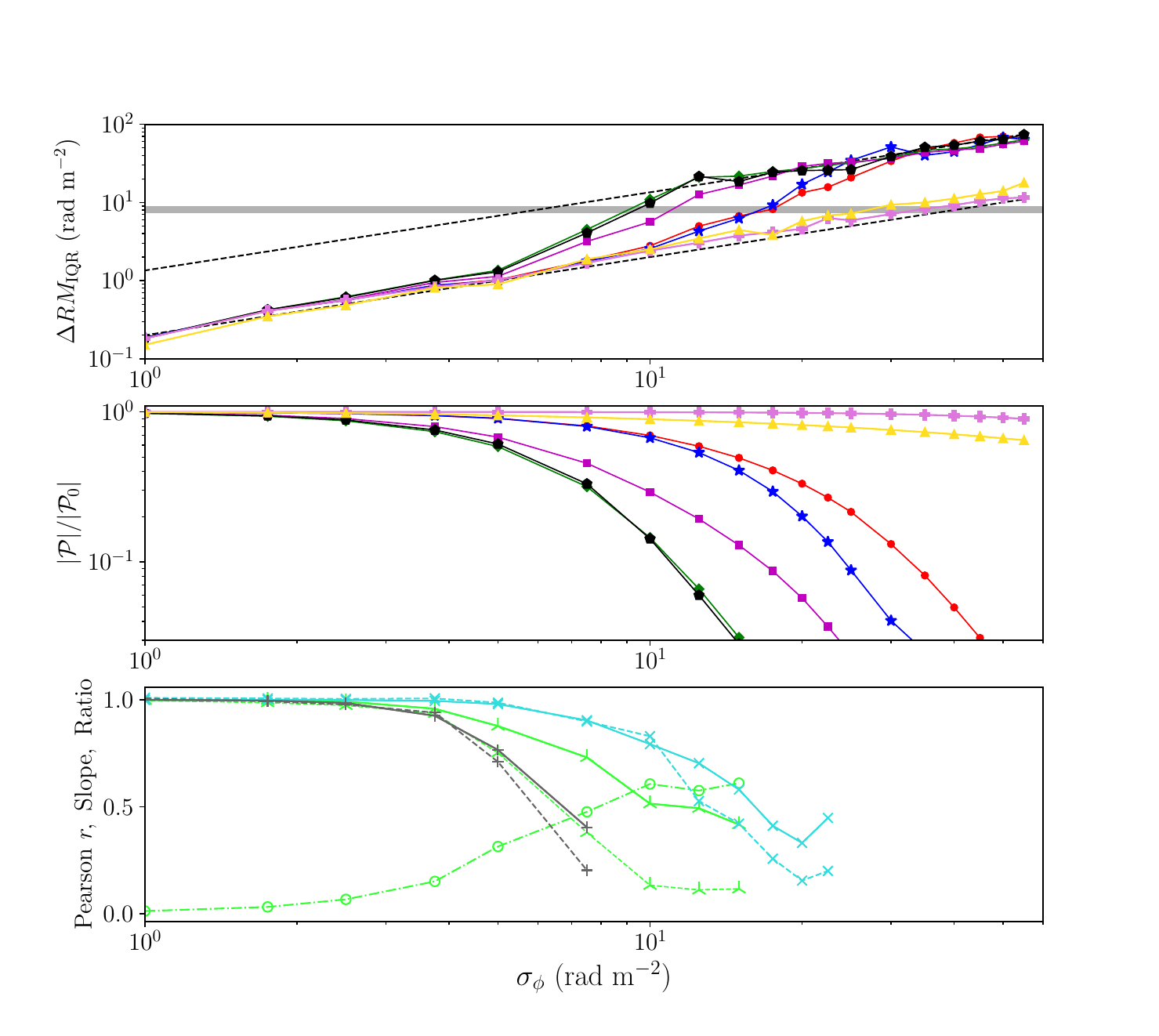}}}
\caption{Top panel: Interquartile range of $RM$ of a sample of sources behind a Faraday screen with power-law Faraday depth structure with index $\gamma = -2.5$, as a function of Faraday depth dispersion across the source $\sigma_\phi$. 
The curves represent surveys listed in Table~\ref{surveys-tab}: POSSUM Low (green curve and diamonds), B001 (black curve and pentagrams), POSSUM Low $+$ Mid (magenta curve and squares), NVSS/CGPS (blue curve and stars), THOR (red curve and dots), Wide (yellow curve and triangles), and B002 (pink curve and crosses). The gray line marks the interquartile range of the 1.4 GHz extragalactic $RM$ dispersion derived from the NVSS $RM$s of \citet{taylor2009} by \citet{schnitzeler2010}, converting standard deviation to IQR assuming a Gaussian distribution. The upper dashed line indicates $\Delta RM_{\rm IQR} = 1.349 \sigma_\phi$, the lower dashed line indicates a convenient reference line, $\Delta RM_{\rm IQR} = \sigma_\phi/5$. Middle panel: median depolarization of the models in the top panel derived from $RM$ synthesis, with the same colors and symbols. Bottom panel: Correlation statistics for a few selected surveys. Solid lines give the Pearson correlation coefficient (green with crow foot: POSSUM-THOR, cyan with $\times$: THOR-B002, grey with $+$: POSSUM-B002). Dashed curves show the slope of the $RM$ correlation for the same surveys. The green dot-dashed curve with circles shows the ratio of the orthogonal residuals of the fit to the POSSUM-THOR correlation to the $RM$ jitter in the THOR survey, as a function of $\sigma_\phi$.
\label{RMjitter-fig}}
\end{figure*}

Figure~\ref{RMsynth-fig} shows results of $RM$ synthesis for a simulated source with $\sigma_\phi = 15\ \radm$ and $\gamma = -2.5$ in six of the surveys listed in Table~\ref{surveys-tab}. This simulation was selected as an outlier $RM$ in the POSSUM Low survey, for the purpose of visualizing important effects. The majority of simulations shows smaller deviations. Each sample of 300 simulated sources has a median $RM$ within statistical errors of the imposed $\phi_0 = 300\ \radm$. The graphs show noiseless, $RM$ Clean deconvolved \citep{heald2009} Faraday depth spectra, all derived from the same spatially integrated model shown on the right. The only difference between the graphs is the wavelength range that was used to perform $RM$ synthesis as defined in Table~\ref{surveys-tab}. 

The thick grey curve represents the histogram of $\phi$ values in the model. The distribution is not Gaussian, and it is subtly different for every simulation. The distribution of the displayed simulation is slightly skewed. This is an important point: random selections of a finite solid angle within the same turbulent screen must result in different Faraday depth distributions. But even a single realization results in significant differences between surveys. The top panel of Figure~\ref{RMsynth-fig} shows $RM$ Clean reconstructed Faraday dispersion functions for THOR (black), NVSS/CGPS (green) and the Wide survey (red). The reconstructed Faraday dispersion functions have different widths because the $RM$ Clean components were restored with a Gaussian that matches the Faraday depth resolution of the survey. The Wide survey approximates the actual Faraday depth distribution well. The THOR and NVSS/CGPS reconstructed Faraday dispersion functions are wider because of the lower Faraday depth resolution. The $RM$s derived from these surveys are within a few $\radm$ from the actual mean of the $\phi$ distribution, which itself is $2.39\ \radm$ different from the imposed sample mean $\phi_0 = 300\ \radm$.  

The bottom panel of Figure~\ref{RMsynth-fig} shows reconstructed Faraday dispersion functions for three low-frequency surveys: POSSUM Low, POSSUM Low $+$ Mid, and B001. These surveys have different but overlapping frequency coverage. The Faraday depths of the peaks are noticeably different. Although it is true that $RM$s from these surveys are correlated, the correlation is not perfect.  Figure~\ref{RMsynth-fig} shows that the reconstructed Faraday dispersion function of a continuous $\phi$ distribution may peak outside the Faraday depth range of the distribution itself, and that even surveys with overlapping frequency coverage may not measure the same $RM$ to within the measurement errors.

\subsection{RM jitter as wavelength-dependent RM scatter}

All curves in Figure~\ref{RMsynth-fig} represent the same simulated source. \textit{The different $RM$s arise from sampling Faraday rotation by the same screen at different wavelengths}. We refer to $RM$ scatter arising from the effects of a turbulent screen as $RM$ jitter, to distinguish it from other sources of $RM$ scatter. Note that the scatter in $\mathcal{R}$ (Equation~\ref{RM-eq}), visualized in Figure~\ref{gradient-fig}, is $RM$ jitter in a narrow band survey.  $RM$ jitter contributes to the total scatter in an $RM$ catalog, but it has peculiar properties. Surveys that measure the same source twice, should find no scatter besides measurement errors. Surveys with different wavelength coverage should find a higher scatter. To understand this phenomenon, we examine the statistics of $RM$ jitter in the simulated samples.

\begin{figure*}
\begin{center}
\resizebox{6cm}{!}{\includegraphics[angle=0,trim={0.0cm 0 0.0cm 0.0cm},clip]{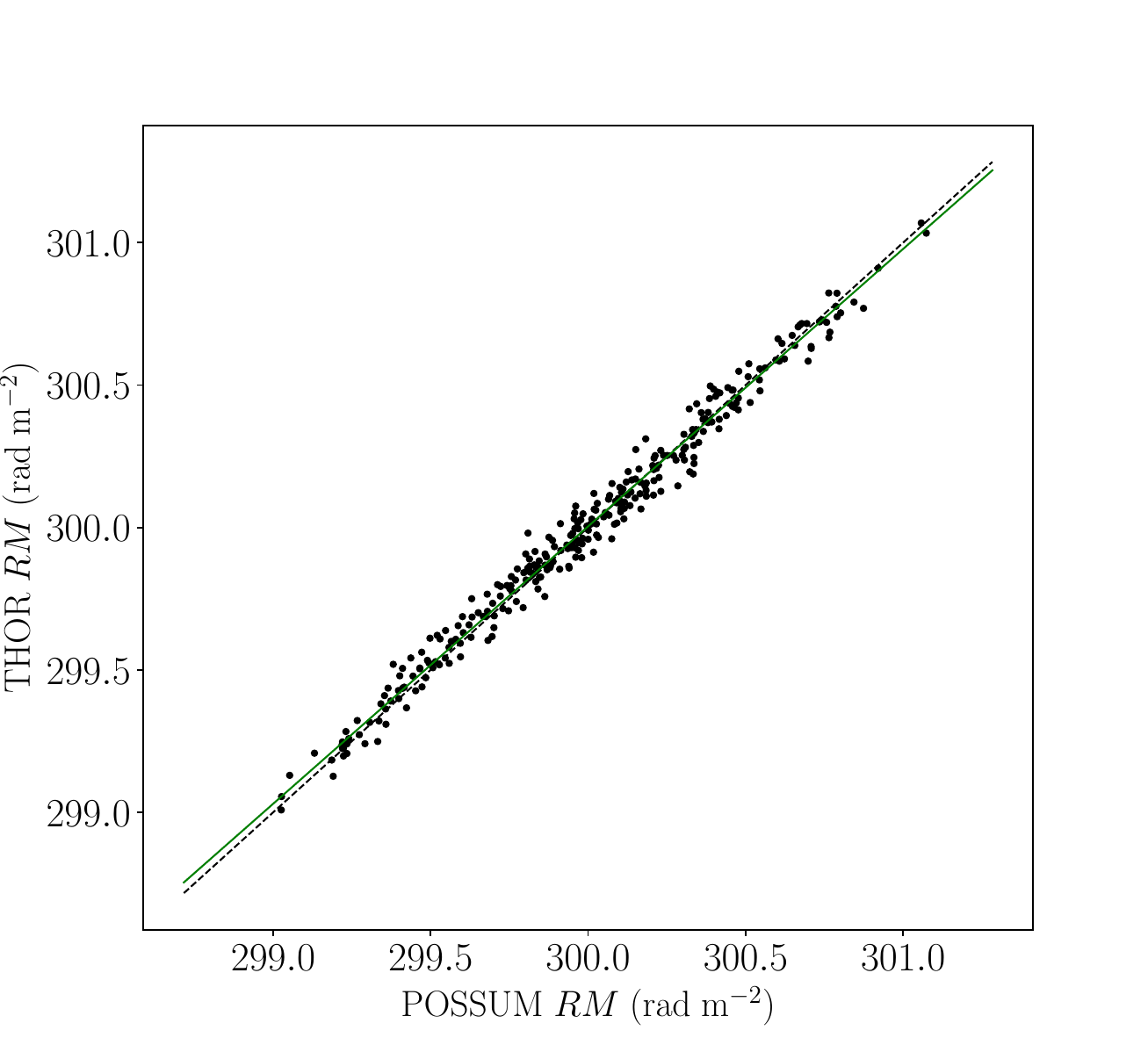}}
\resizebox{6cm}{!}{\includegraphics[angle=0,trim={0.0cm 0 0.0cm 0.0cm},clip]{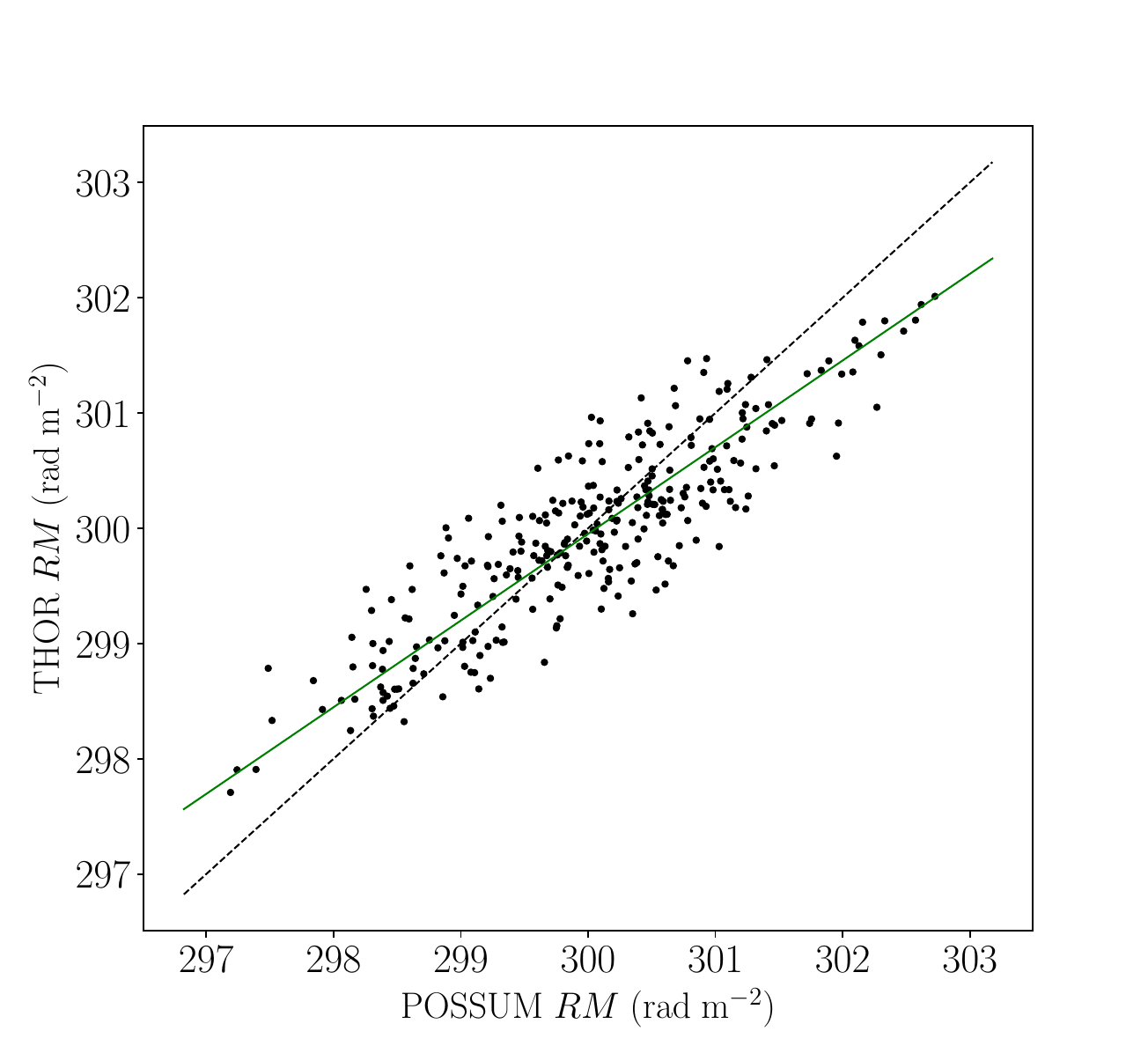}}\\
\resizebox{6cm}{!}{\includegraphics[angle=0,trim={0.0cm 0 0.0cm 0.0cm},clip]{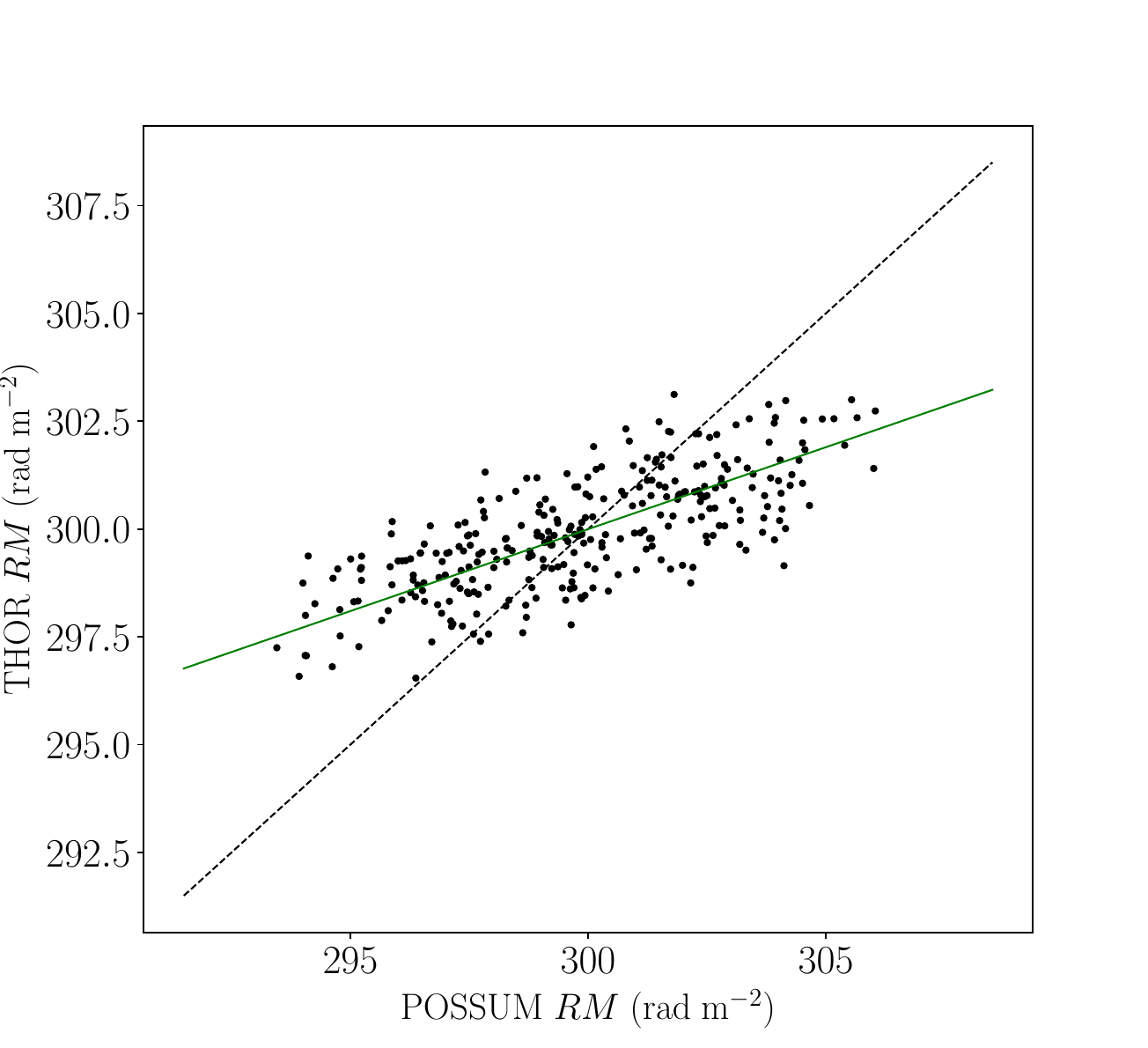}}
\resizebox{6cm}{!}{\includegraphics[angle=0,trim={0.0cm 0 0.0cm 0.0cm},clip]{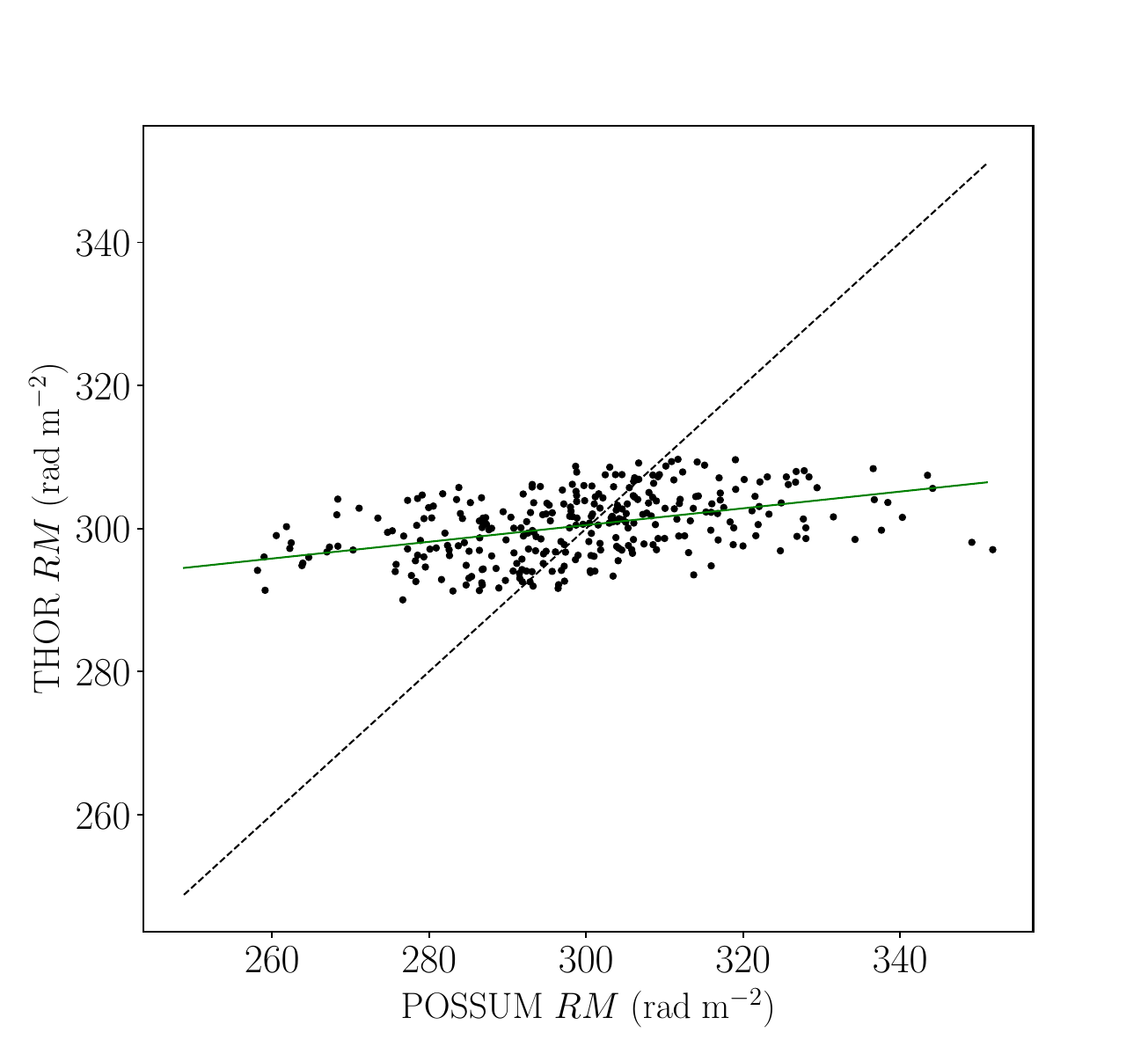}}
\end{center}
\caption{ Correlation between $RM$ from the surveys POSSUM Low and THOR for $\gamma=-2.5$ and four values of $\sigma_\phi$.  Top left panel: $\sigma_\phi = 2.5\ \radm$ with $r = 0.992$ and slope 0.966, Top right panel: $\sigma_\phi = 5\ \radm$ with $r = 0.877$ and slope 0.683, Bottom left panel: $\sigma_\phi = 7.5\ \radm$ with $r = 0.731$ and slope 0.345, Bottom right panel: $\sigma_\phi = 15\ \radm$ with $r = 0.417$ and slope 0.110. The black dashed lines indicate equal $RM$. The solid green lines show the fit as described in the text.   
\label{RMcorr_2-fig}}
\end{figure*}

Results of the simulations with $\gamma=-2.5$ are summarized in Figure~\ref{RMjitter-fig}. The top panel shows the interquartile range (IQR) of the $RM$ distributions of samples of 300 simulated sources, each with the same combination of $\gamma$ and $\sigma_\phi$, but independent realizations of the screen for each source. Samples with different $\sigma_\phi$ are all mutually independent (no rescaling of simulated images). 

For small $\sigma_\phi$, the $RM$ jitter in all surveys increases approximately as $\Delta RM_{\rm IQR} \approx \sigma_\phi/5$, indicated by the lower dashed line in Figure~\ref{RMjitter-fig}. We refer to this as the linear regime of $RM$ jitter. The linear rise in $RM$ jitter is also apparent along the line that marks the frequency $2\ \rm GHz$ in Figure~\ref{gradient-fig}, where $\lambda^2 < 1/\sigma_\phi$. We will see that strength of $RM$ jitter in the linear regime also depends on $\gamma$. 

As $\sigma_\phi$ increases, the low-frequency surveys are the first to display a rapid increase in $RM$ jitter. We refer to this as the non-linear regime. For high-frequency surveys, the non-linear regime occurs for sources with a higher $\sigma_\phi$. 

At even higher values of $\sigma_\phi$, the curves converge again, approximately to the line that defines the IQR of a Gaussian distribution with standard deviation $\sigma_\phi$, which is $\Delta RM_{\rm IQR} = 1.349\, \sigma_\phi$, indicated by the upper dashed line in Figure~\ref{RMjitter-fig}. We refer to this as the saturated regime. We will see that $RM$ jitter in the saturated regime does not depend on $\gamma$. 

$RM$ jitter in Figure~\ref{RMjitter-fig} is related to the depolarization factor $|\mathcal{P}|/|\mathcal{P}_0|$ shown in the middle panel of Figure~\ref{RMjitter-fig}. The POSSUM Low survey and the B001 survey show the same $RM$ jitter and depolarization, even though it is clear from Figure~\ref{RMsynth-fig} that $RM$s from individual sources are not necessarily perfectly correlated. These surveys have the same sensitivity to continuous Faraday depth structure, $W_{\rm max}$. The Faraday depth resolution of the B001 survey is a factor $\sim 3$ worse, showing that $RM$ jitter is not an effect of resolving the actual Faraday depth distribution. Sources that suffer more depolarization will be under-represented in a flux limited sample. $RM$ jitter for a flux-limited sample will be discussed Section~\ref{flux_limited-sec}.
 
\begin{figure*}
\centerline{\resizebox{12.0cm}{!}{\includegraphics[angle=0,trim={0.0cm 1.0cm 0.0cm 2.5cm}]{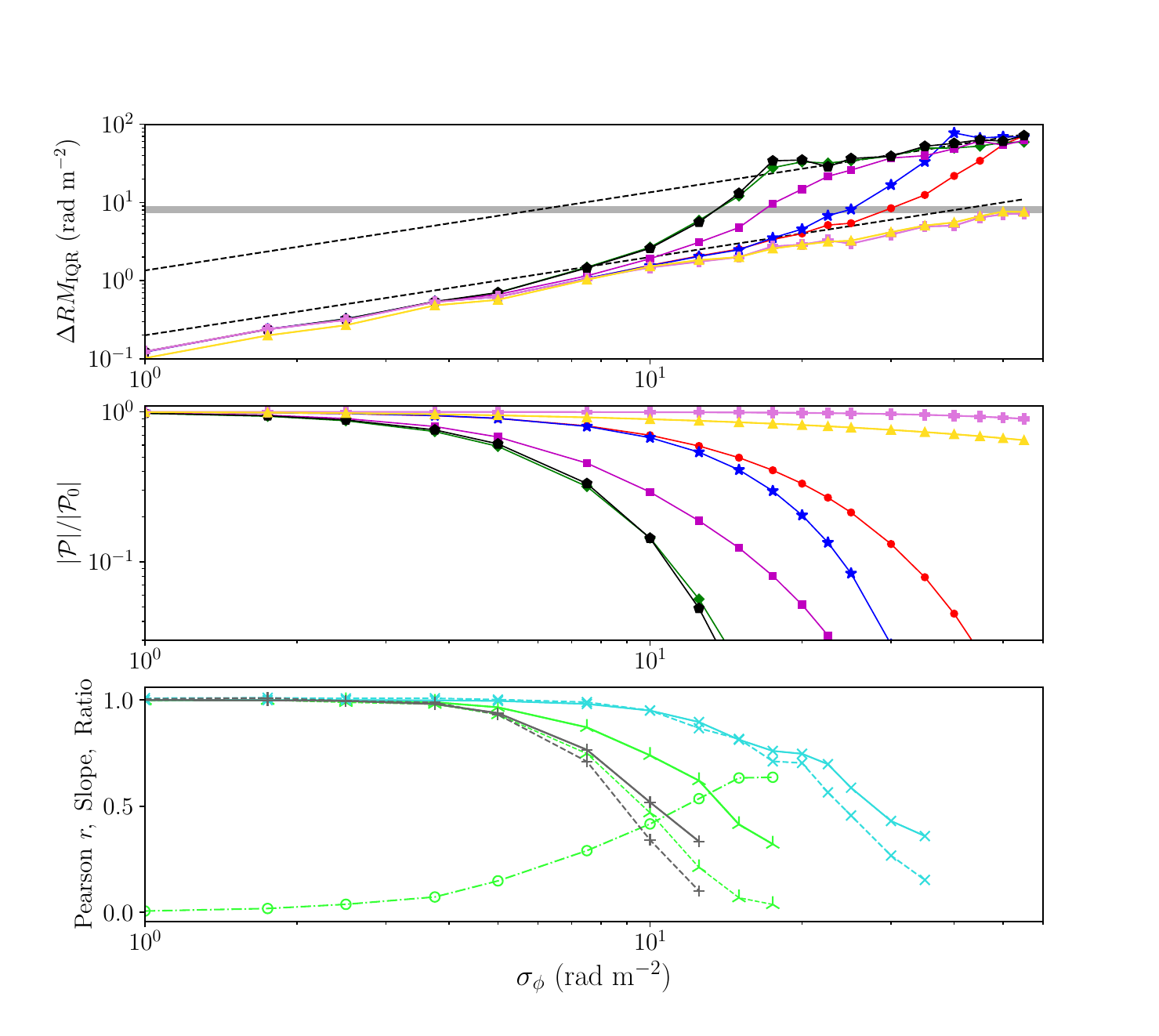}}}
\caption{As Figure~\ref{RMjitter-fig}, for $\gamma=-2.0$. The linear, non-linear and saturated regimes exist at higher $\sigma_\phi$. Note that $\Delta RM_{\rm IQR}$ in the linear regime drops below the reference line $\Delta RM_{\rm IQR} = \sigma_\phi/5$, indicating reduced $RM$ jitter because there is more power on small scales. $RM$ jitter in the saturated regime approaches the same line as in Figure~\ref{RMjitter-fig}.\label{RMjitter2-fig}}
\end{figure*}

\begin{figure*}
\centerline{\resizebox{12.0cm}{!}{\includegraphics[angle=0,trim={0.0cm 1.0cm 0.0cm 2.5cm}]{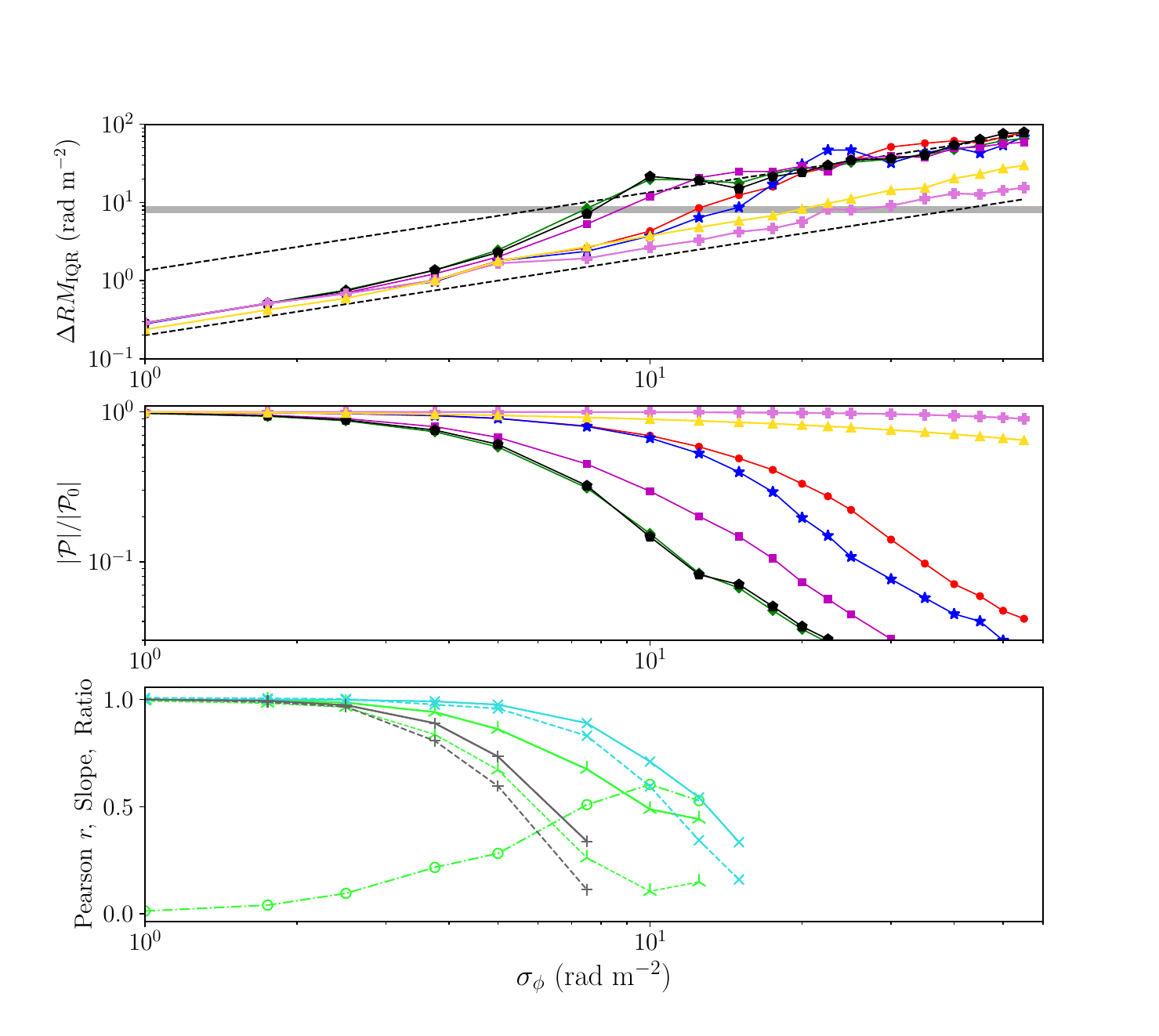}}}
\caption{As Figure~\ref{RMjitter-fig}, for $\gamma=-3.0$. The linear, non-linear and saturated regimes exist at lower $\sigma_\phi$. Note that $\Delta RM_{\rm IQR}$ in the linear regime is above the reference line $\Delta RM_{\rm IQR} = \sigma_\phi/5$, indicating increased $RM$ jitter because there is more power on large scales. $RM$ jitter in the saturated regime approaches the same line as in Figure~\ref{RMjitter-fig}. \label{RMjitter3-fig}}
\end{figure*}

When comparing $RM$s of the same source from different surveys, a certain degree of correlation is expected. The Pearson $r$ correlation coefficient between a few representative surveys is shown as a function of $\sigma_\phi$ in the bottom panel of Figure~\ref{RMjitter-fig}. Correlations of $RM$ from POSSUM Low and $RM$ from THOR for four samples with different $\sigma_\phi$ are shown in Figure~\ref{RMcorr_2-fig}. 

In the linear regime, the correlation between $RM$s of identical sources observed by different surveys is near perfect. As $\sigma_\phi$ increases, the correlation deteriorates, as shown by a decreasing Pearson $r$ correlation coefficient in the bottom panel of Figure~\ref{RMjitter-fig}, and in Figure~\ref{RMcorr_2-fig}. Interestingly, the correlation between POSSUM and THOR remains linear but its slope decreases as $\sigma_\phi$ increases, as if the $RM$ jitter is enlarged by a multiplicative factor that depends on depolarization. Indeed, the slope of the correlations is to a good approximation the ratio of the depolarization factors of the two surveys for the assumed value of $\sigma_\phi$. For example, the slope of the correlation in the top right panel of Figure~\ref{RMcorr_2-fig} is 0.683, whereas the ratio of the median depolarization factors is 0.650. The slope in the bottom left panel of Figure~\ref{RMcorr_2-fig} is 0.345 whereas the ratio of the median depolarization factors is 0.397. In both cases the formal error in the fitted slope is 0.020, so the differences with the depolarization ratios are barely significant. When comparing $RM$s from surveys with overlapping wavelength coverage, such as POSSUM and B001, the correlation between $RM$s is generally tighter, but decorrelation does occur for larger $\sigma_\phi$ related specifically to the subset of sources with the strongest depolarization within the observed wavelength range.  

This suggests that $\Delta RM_{\rm IQR}$ enters the non-linear regime as depolarization sets in, because of $1/|\mathcal{P}|$ amplification of differential Faraday rotation in the $QU$ plane. Multiplication of the $RM$ of every source with its individual depolarization factor rectifies the slopes of the correlations in Figure~\ref{RMcorr_2-fig}, but it does not remove the scatter. Amplification of $RM$ associated with depolarization is not new. \citet{anderson2016} (their Figure 5 and Figure 16) show sources with large swings in $\mathcal{R}$ associated with depolarization in certain wavelength ranges. This is also the effect that \citet{gaensler2011} aimed to remove from their analysis when they introduced the concept of polarization gradients that are invariant under translation in the $QU$ plane.

Figure~\ref{RMjitter2-fig} and Figure~\ref{RMjitter3-fig} illustrate the effect of the slope of the power spectrum, $\gamma$. In case of a flatter power spectrum (Figure~\ref{RMjitter2-fig}), $\Delta RM_{\rm IQR}$ in the linear regime is reduced, whereas it is increased in the case of a steeper power spectrum (Figure~\ref{RMjitter3-fig}), for the same $\sigma_\phi$. This relates $RM$ jitter to Faraday depth structure on larger scales within the beam. However, $\Delta RM_{\rm IQR}$ in the saturated regime approaches the same line, corresponding to the IQR of a Gaussian with dispersion $\sigma_\phi$, in all three cases. The saturated regime is reached for lower $\sigma_\phi$ if the power spectrum is steeper. This has a significant consequence: depolarization in the saturated regime for $\gamma = -2.0$ is so strong that we do not expect to observe those sources. However, depolarization in the saturated regime is much less when $\gamma = -3.0$, suggesting that some of those sources may be observed.

The dependence of depolarization on $\sigma_\phi$ also changes with the slope of the power spectrum, $\gamma$. When the power spectrum is less steep (Figure~\ref{RMjitter2-fig}), the depolarization graph curves down as the exponential model in Equation~\ref{depol_model-eq} would. When the power spectrum is steeper, the depolarization graph approximates a power law in $\sigma_\phi$. This is analogous to the median depolarization law derived by \citet{tribble1991}. The monochromatic case considered by \citet{tribble1991} scales according to the dimensionless variable $\sigma_\phi \lambda^2$, so this result is equivalent to the result of \citet{tribble1991} for $|\mathcal{P}(\lambda^2)|/|\mathcal{P}_0|$. In case of broad band surveys, the scaling $\sigma_\phi \lambda_0^2$ is not exact, as can be seen by the difference between the depolarization curves of THOR and NVSS/CGPS. Since the depolarization also depends on $\gamma$, the slope of the power spectrum is a confounding parameter when modelling depolarization by a turbulent plasma. 

The simulations show a modest over-shoot at the onset of the saturated regime that seems to coincide with an inflection point in the depolarization curves, especially in Figure~\ref{RMjitter3-fig}. Investigation of this higher-order effect is deferred to future work.

An important detail is the difference between broad-band surveys and narrow band surveys. This is relevant for the difference between the monochromatic $\mathcal{R}$ and the rotation measure $RM$. For narrow band surveys, $RM$ derived from a fit of $\theta(\lambda^2)$ or from $RM$ synthesis, is approximately equal to $\mathcal{R}$. If $\mathcal{R}$ is constant over a wider frequency range, a wide-band survey using $RM$ synthesis will also find $RM = \mathcal{R}$. Wavelength-dependent depolarization by a turbulent screen can impose significant wavelength dependence on $\mathcal{R}$, in particular near a minimum of $|\mathcal{P}|$, as is the case for some analytic models \citep[e.g.][]{burn1966,sokoloff1998}. A narrow-band survey would report whatever the value of $\mathcal{R}$ is in the observed wavelength range. A broad-band survey will (usually) report one $RM$. The wide-band $RM$ need not have the same sign as the $RM$ from the narrow band survey.
 
 \begin{figure*}
\resizebox{10cm}{!}{\includegraphics[angle=0,trim={1.3cm 0 2.0cm 2.8cm},clip]{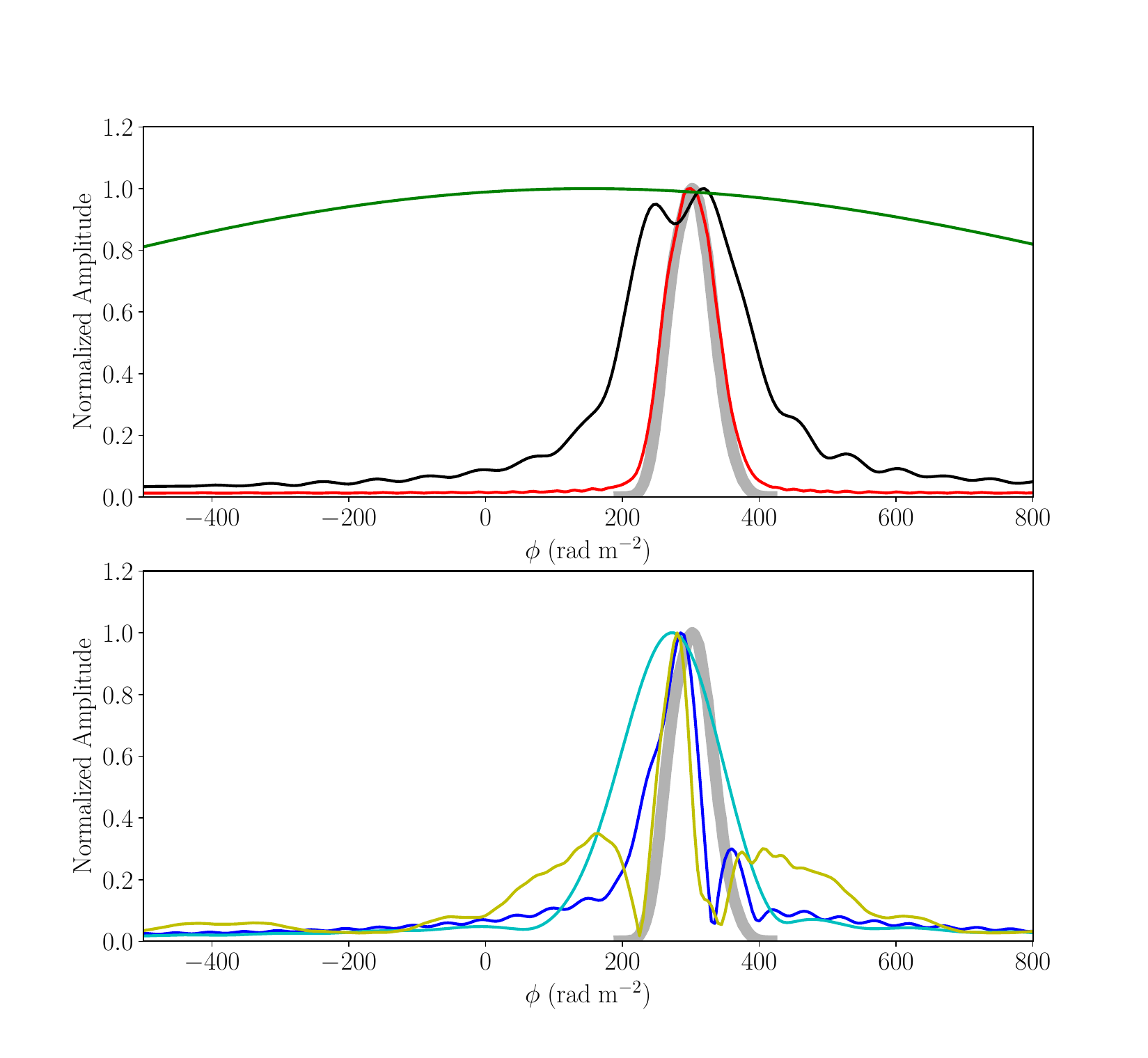}}
\resizebox{6cm}{!}{\includegraphics[angle=0,trim={3.5cm 0 1.5cm 3.8cm},clip]{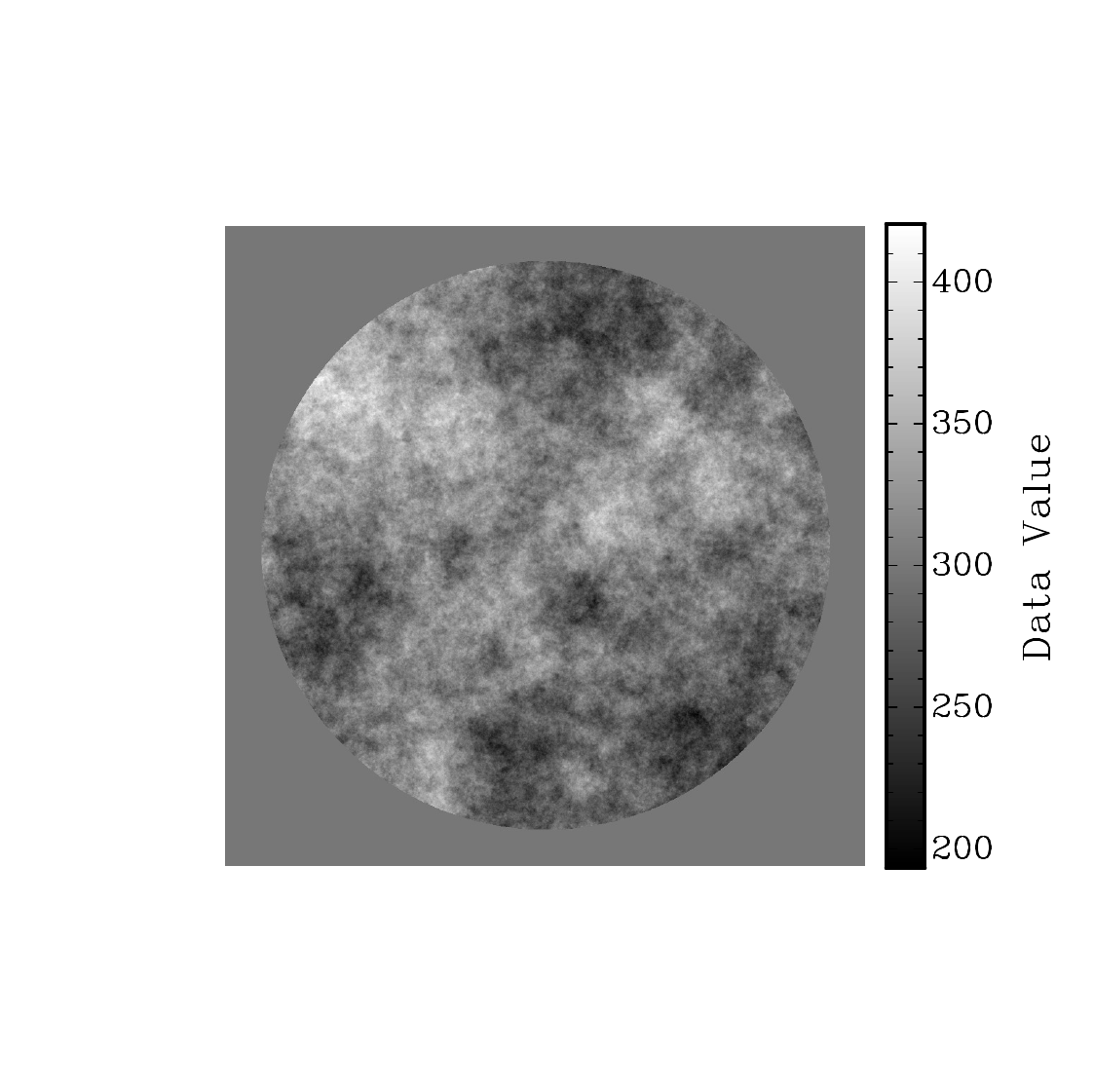}}
\caption{Deconvolved Faraday dispersion function for 6 surveys of a single simulated source with Faraday depth distribution as shown on the right, with $\sigma_\phi=30\ \radm$ and $\gamma=-2.5$. Note that the range of the Faraday depth axis is almost twice the range shown in Figure~\ref{RMsynth-fig}. The histogram of the model Faraday depths is drawn as a thick gray curve, with full range $194.02\ \radm$ to $419.71\ \radm$. Top panel: THOR survey (black, $RM=318.35\ \radm$), NVSS/CGPS survey (green, $RM=151.10\ \radm$), and full simulated wavelength range of 600 MHz to 8 GHz (red, $RM = 298.26\ \radm)$. This simulation was selected as a more extreme deviation for the NVSS/CGPS survey for demonstration purposes. The NVSS/CGPS curve and the THOR curves are not representative for the complete sample.  Bottom panel: POSSUM Low (blue, $RM=286.50\ \radm$), POSSUM Low $+$ Mid (yellow, $RM=280.40\ \radm$), B001 (cyan, $RM=272.55\ \radm$). \label{RMsynth2-fig}}
\end{figure*}

Some models  indeed show extreme values of $\mathcal{R}$ in narrow wavelength intervals, leading to extreme scatter including reversals of the sign of $RM$ even though $\phi_0 = 300\ \radm$. This is why the IQR is preferred to quantify $RM$ jitter over the standard deviation, in particular for the narrow band surveys. Figure~\ref{RMsynth2-fig} shows $RM$ synthesis for a simulated source with $\sigma_\phi = 30\ \radm$ that was selected as an extreme outlier $RM$ in the NVSS/CGPS survey that arises from an anomalous value of $\mathcal{R}$ in the narrow band survey related to partial depolarization. (see also Figure~\ref{gradient-fig}). The effect of the localized extreme $\mathcal{R}$ within the wider L-band is unusual Faraday complexity in the THOR survey, where the reconstructed Faraday dispersion function shows a double peak that can be misinterpreted as a superposition of two discrete components. A system of two Faraday thin components can also depolarize and repolarize over a certain wavelength range. It should not be surprising that a turbulent plasma that depolarizes and repolarizes within the observed band resembles a two-component source. Such extremes in $\mathcal{R}$ occur only in $\sim 1\%$ of simulated sources with significant depolarization, so the fraction of sources in a survey that displays a certain form of complexity can be an effective way to discriminate between competing models.

A consequence of $RM$ jitter is that narrow band surveys must contain sources with extreme $RM$ that would be erroneously attributed to unexplained errors when compared with broad-band data. Although the narrow band $RM$ may accurately describe Faraday rotation of the source within the observed $\lambda^2$ range, the reported $RM$ can easily be misinterpreted. 

The main take-away is that $RM$ scatter in a survey depends on the amount of Faraday dispersion in the source, the power spectrum of turbulent structure, as well as the wavelength coverage of the survey. This complicates drawing conclusions about the turbulent properties of a plasma or the existence of large-scale magnetic fields.

Somewhat counter-intuitive, $RM$ jitter is not an effect of resolving the source in Faraday depth. $RM$ jitter in the POSSUM Low survey and the B001 survey is statistically the same, while the B001 survey has three times coarser resolution in Faraday depth. $RM$ jitter in POSSUM Low $+$ Mid is less than in POSSUM Low.  Also, $RM$ jitter in POSSUM Low, B001, and POSSUM Low $+$ Mid is stronger than in the narrow-band surveys NVSS/CGPS, but $RM$ jitter in the narrow band surveys exceeds that of the broad L-band survey THOR. $RM$ jitter in the Wide survey remains low over the $\sigma_\phi$ range under investigation, but $RM$ jitter in the B002 survey with a reduced bandwidth is even smaller. $RM$ jitter is independent of Faraday depth resolution because it is an effect of the Faraday depth distribution within the source and depolarization by differential Faraday rotation, both of which are independent of Faraday depth resolution. 

The contribution of $RM$ jitter to the total scatter in $RM$ should also be considered in the context of $RM$ errors. Measurement errors at fixed signal to noise ratio are determined by Faraday depth resolution.  Table~\ref{surveys-tab} lists the $RM$ errors for a 10$\sigma$ detection in polarized flux density. While the potential for outliers is greatest in narrow band surveys, $RM$ jitter is much more significant compared with $RM$ errors in broadband surveys with smaller $RM$ errors. 

The results presented here should not give the impression that high-frequency surveys are immune to non-linear $RM$ jitter. This is only the case for the $\sigma_\phi$ range investigated here. The simulations scale approximately as $\sigma_\phi \lambda_0^2$, with some differences related to averaging over a wide frequency range. Surveys at high frequencies can detect sources with higher $\sigma_\phi$ that are known to exist, and have correspondingly high $RM$ jitter. $RM$ jitter is less in the B002 survey than in the Wide survey because it lacks the longer wavelengths at which depolarization occurs. This apparent advantage is negated by the reduced resolution in Faraday depth, which increases measurement errors (see Equation~\ref{RM_error-eq} and Table~\ref{surveys-tab}).

\subsection{RM jitter in a flux density limited sample}
\label{flux_limited-sec}
\begin{figure*}
\centerline{\resizebox{12cm}{!}{\includegraphics[angle=0,trim={0.0cm 1.0cm 0.0cm 2.5cm}]{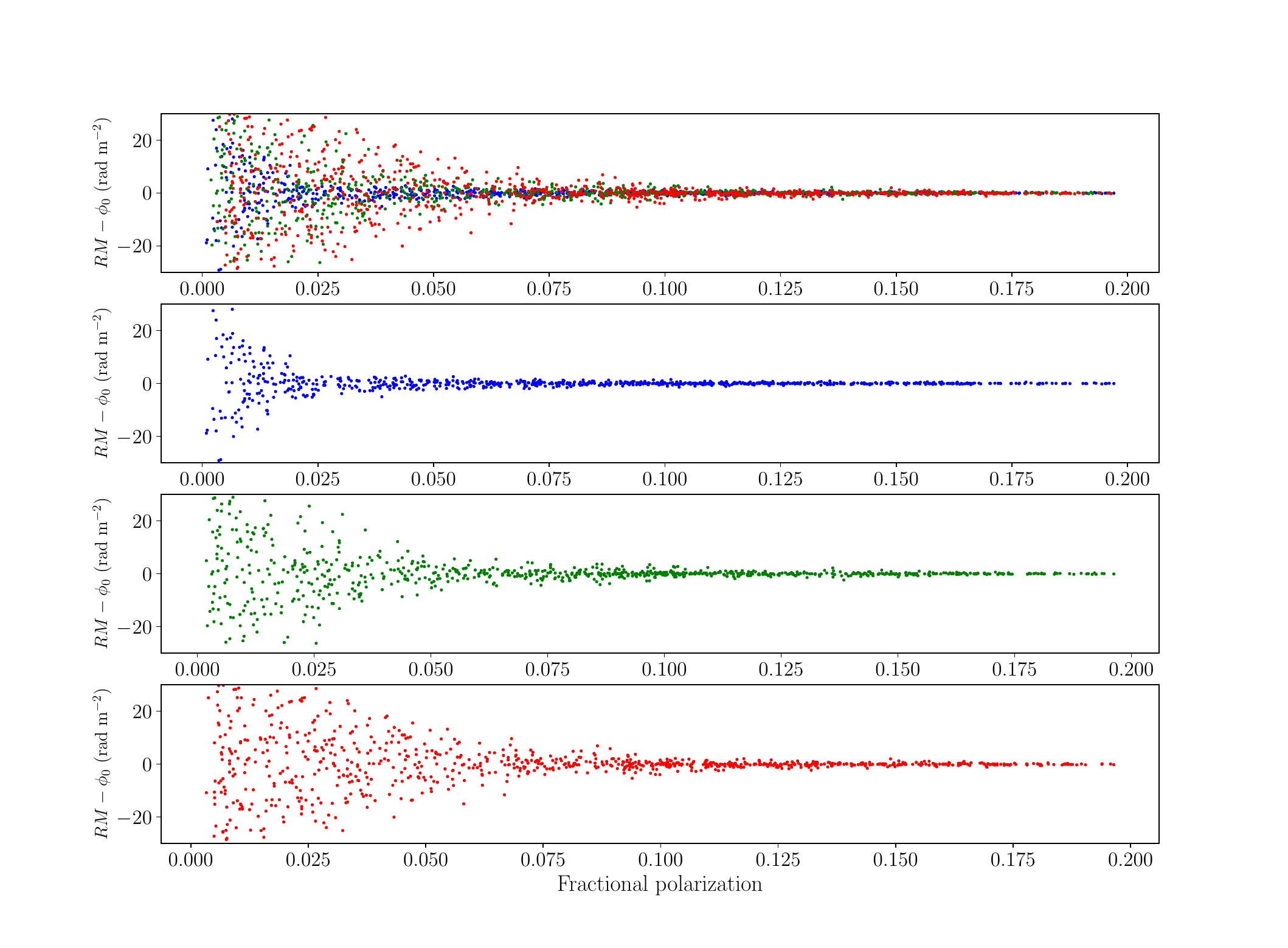}}}
\caption{RM jitter as a function of fractional polarization in a simulated sample with polarized flux density threshold as described in the text. The top panel shows the combined sample with $\gamma = -2.0$ (blue),  $\gamma = -2.5$ (green), and $\gamma = -3.0$ (red). The subsequent panels show the three subsamples separately. Note that the slope of the power spectrum, $\gamma$, is an unknowable parameter for an unresolved radio source. \label{fracpol_jitter-fig}}
\end{figure*}

An $RM$ grid is a collection of $RM$s from a survey of radio sources \citep{beck2004}. This is usually a polarized flux density limited sample. Sources with a higher $\sigma_\phi$ are under-represented, as only the brighter sources with strong depolarization (stronger $RM$ jitter) may be detectable. Each survey will have its own sensitivity to a range of $\sigma_\phi$ as indicated by $W_{\rm max}$. Selection effects related to sensitivity favour sources in the linear or slightly non-linear regime of $RM$ jitter that suffer negligible or modest wavelength-dependent depolarization (Figure~\ref{RMjitter-fig}), although some sources near the saturation regime are detectable, \citep[see e.g.][for a discussion]{shanahan2023}. This is also apparent from the range of $\sigma_\phi$ detected in extragalactic sources in the wavelength range considered here \citep{osullivan2017,livingston2021}.

We assess $RM$ jitter in a flux-density limited sample for the POSSUM Low $+$ Mid survey, motivated by a recent result of \citet{vanderwoude2024}, who reported that $RM$ scatter depends on fractional polarization. For sources that are less than $3\%$ polarized, $\Delta RM_{\rm IQR} \sim 15\ \radm$, while for sources that are more than 3\% polarized, $\Delta RM_{\rm IQR} \sim 6\ \radm$. Previously, \citet{bernet2012} had noticed a higher $RM$ scatter for sources in their sample that were less than 3.2\% polarized at 21 cm. \citet{vernstrom2019}, their Figure 8, showed a steep increase in $RM$ scatter toward lower fractional polarization in their sample of random (unrelated) pairs of radio sources. While these results are in approximate agreement with \citet{vanderwoude2024}, the latter authors made the significant observation that the effect is independent of the signal to noise ratio of the polarized signal.

A few assumptions are required to calculate $RM$ jitter in a flux limited sample. The simulations provide a dimensionless wavelength-dependent depolarization factor and an $RM$ for each source. The polarized flux density is calculated using a total flux density $S$ and fractional polarization $\Pi$. A simulated source list of total flux densities was created by random draws from a fit to the 1.4 GHz radio source counts by \citet{windhorst2003} in the range  $1 < S_{1.4} <  300$ mJy. A minor difference in the frequency of these source counts and the average frequency of the source catalog of \citet{vanderwoude2024} is not important because the slope of the differential source counts will not be noticeably different. Each source is assigned a $\sigma_\phi$ drawn from an assumed distribution. The actual distribution of $\sigma_\phi$ is not known. Half-Gaussian and uniform distributions were tried. Since there are lobes of radio galaxies with Faraday depth range much larger than the $\sigma_\phi$ considered here, a uniform distribution of $\sigma_\phi$ was preferred. The $\sigma_\phi$ distribution of the sample then results from selection effects arising from the detection threshold in polarization and depolarization.

The fractional polarization may be calculated from the depolarization factor and the intrinsic fractional polarization.  The theoretical maximum fractional polarization of incoherent synchrotron radiation in a uniform magnetic field is $\sim 70\%$. High-resolution observations of nearby radio galaxies suggest the intrinsic fractional polarization of radio lobes to be in the range of 10\% to 30\%, sometimes higher \citep{laing2008,osullivan2013,anderson2018,baidoo2023}. The calculated $RM$ jitter will correlate with the adopted value, but may be further constrained by the median fractional polarization of the complete sample. Experimentation showed that the higher end of fractional polarization drives up $\Delta RM_{\rm IQR}$ because it allows sources with stronger depolarization factors into the sample. The intrinsic fractional polarization was drawn from a uniform distribution between $10\%$ and $20\%$. The intrinsic fractional polarization was multiplied by the depolarization factor of sources from the POSSUM Low $+$ Mid survey, for comparison with the results from \citet{vanderwoude2024}. Any source with depolarization factor less than 0.01 was assigned a fractional polarization equal to zero. A detection threshold of $0.1\ \rm mJy$ in polarized flux density was applied following \citet{vanderwoude2024}. For the evaluation of the sample fractional polarization, the sample was further restricted to total flux density more than 10 mJy. This ensures that sources with $1\%$ fractional polarization are above the detection threshold in polarized flux density \citep[see][]{vanderwoude2024}.

\begin{figure*}
\centerline{\resizebox{10cm}{!}{\includegraphics[angle=0,trim={0.0cm 0.5cm 0.0cm 0.5cm}]{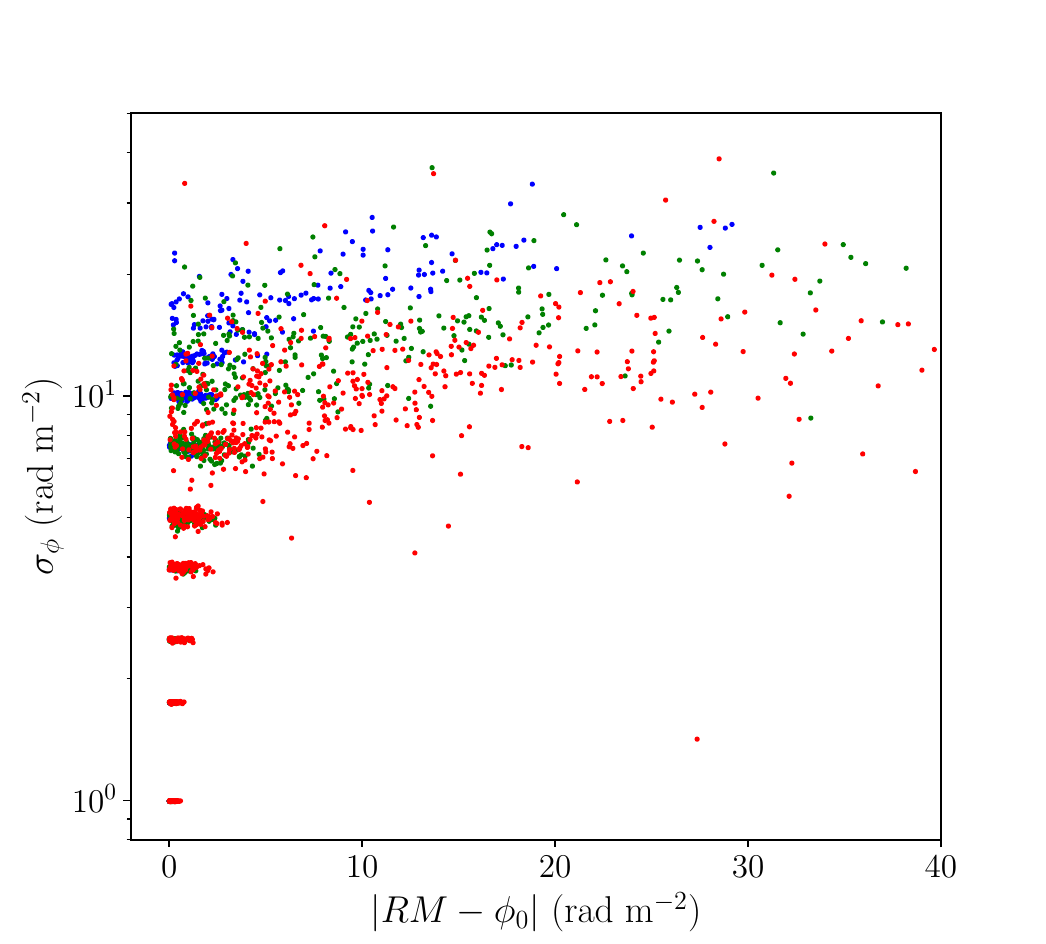}}}
\caption{Faraday dispersion $\sigma_\phi$ as a function of $RM$ for a simulated sample of radio sources as described in the text, derived from fitting Equation~\ref{depol_model-eq}. The simulated survey is POSSUM Low + Mid. No noise was added to the models. When $\sigma_\phi$ is small, the fits recover the Faraday dispersion of the models. For larger $\sigma_\phi$, the scatter increases in an absolute sense and as a fraction of the fitted Faraday dispersion, as is evident from the increasing scatter along the logarithmic $\sigma_\phi$ axis. The reason for this is the stochastic and non-Gaussian depolarization of the models. Colours of the symbols follow the definitions in Figure~\ref{fracpol_jitter-fig}. \label{sigma_jitter-fig}}
\end{figure*}

The expected $RM$ jitter depends on the adopted slope of the power spectrum, $\gamma$. When comparing the top two panels of Figures~\ref{RMjitter-fig}, \ref{RMjitter2-fig}, and \ref{RMjitter3-fig}, it becomes apparent that the linear regime of the simulations with $\gamma = -2.0$ extends to higher $\sigma_\phi$ (stronger depolarization) than in the case of $\gamma = -2.5$ and $\gamma = -3.0$. Since there is more power on smaller angular scales when $\gamma = -2.0$, this simulation is closer to the limit of a large number of independent cells within the beam that is the foundation of Equation~\ref{depol_model-eq}. In this limit, we expect the $RM$ jitter to be zero.

Experiments with samples that assume a single value of $\gamma$ qualitatively show a relation between $RM$ scatter and fractional polarization, $\Pi$. Simulations that assume $\gamma = -2.0$ under-estimate the observed scatter for sources in POSSUM Low $+$ Mid with $\Pi < 3\%$ by a factor $\sim 2$, as shown in Figure~\ref{fracpol_jitter-fig}. Simulations that assume $\gamma = -3.0$ over-estimate this scatter by $\sim 50\%$. Simulations that assume $\gamma = -2.5$ come close, although they perform better when a half-Gaussian distribution reduces the number of sources with a large $\sigma_\phi$. This suggests that a single value $\gamma \gtrsim -2.5 $ could reproduce the observed $RM$ scatter in POSSUM Low $+$ Mid for sources with $\Pi < 3\%$. Combining all three simulated values of $\gamma$, with equal probability, results in $\Delta RM_{\rm IQR} = 18.3\ \radm$ for the sample with $\Pi < 3\%$. The simulated sample with $\Pi > 3\%$ has $\Delta RM_{\rm IQR} = 0.94\ \radm$. This is well below the $6\ \radm$ found by \citet{vanderwoude2024}, indicating that $RM$ scatter from other effects dominates over $RM$ jitter for this subsample.  The median fractional polarization of all sources brighter than 10 mJy in the simulated sample is $2.0\%$, which is close to the observed value \citep[e.g.][]{stil2014,rudnick2014,vanderwoude2024}.

These results are encouraging, indicating that a distribution of $\gamma$ that is modestly favouring a power spectrum that on average is slightly flatter than $\gamma = -2.5$ could reproduce the result precisely. At this time, available observational data for $\gamma$ are extremely limited. The significance of the slope of the power spectrum of Faraday depth structure to depolarization and $RM$ scatter is noteworthy, as it relates an important parameter of the turbulence in the intergalactic plasma to observable quantities. 

\citet{osullivan2023} found a robust standard deviation of $2\ \radm$, after correcting for Galactic Faraday rotation, for their $RM$ grid made with the Low Frequency Array (LOFAR) in the frequency range $120\ \rm MHz$ to $168\ \rm MHz$. For a Gaussian distribution, this translates to $\Delta RM_{\rm IQR} = 2.70\ \radm$. This is smaller than the range found for POSSUM by \citet{vanderwoude2024}. However, $W_{\rm max}$ is only $0.32\ \radm$ for the LOFAR survey. The $RM$ jitter is expected to be quite small, as it would saturate at $\sigma_\phi \sim 0.32$ for an inter-quartile range of $1.39\sigma_\phi = 0.43\ \radm$. This suggests that also in the case of \citet{osullivan2023}, the observed $RM$ scatter is dominated by other effects than $RM$ jitter.

A minimum threshold in fractional polarization is an imperfect but practical tool to reduce $RM$ scatter in an observed sample of radio sources. $RM$ jitter is directly related to Faraday complexity. However, there are practical challenges in pursuing measures of Faraday complexity as a means to reduce $RM$ scatter in a survey. Measuring Faraday complexity is close to impossible near the detection limit of a survey, whereas fractional polarization is always available. Quantifying Faraday complexity is notoriously complicated \citep[e.g.][]{sun2015}. Figure~\ref{sigma_jitter-fig} shows results of $QU$ fitting of the model in Equation~\ref{depol_model-eq} to a simulated sample of sources with no noise. Selecting sources with small Faraday depth dispersion could reduce the effects of $RM$ jitter in the sample, but only at the expense of a significant fraction of the sample. This will be worse for sources affected by noise.

\section{A system of $N$ Faraday thin components}
\label{Ncomp-sec}

An informative illustration of $RM$ jitter is a sample of sources, each with $N$ Faraday thin components that occupy a range of Faraday depth that is not resolved by a survey. The effect of relative polarization angle on $RM$ in the case $N=2$ was discussed previously by \citet{kumazaki2014}. In the context of this paper, this case shows that differential Faraday rotation between two Faraday depth components can lead to wavelength dependence of $\mathcal{R}$. For arbitrary $N$, the complex linear polarization is
\begin{equation}
\mathcal{P}(\lambda^2) = \sum_{j = 1}^{N} |\mathcal{P}_{0,j}| \exp[2i (\theta_{0,j} + \phi_j \lambda^2)].
\label{Ncomp-eq}
\end{equation}
This is the discrete equivalent of the integral in Equation~\ref{RMsynth_inv-eq}. The mean of the $\phi_j$ appears only as a constant complex exponential that can be pulled out of the sum as a trivial rotation, so we assume that the expectation value, which is the unknowable true mean, of the $\phi_j$ is zero. The mean of the $\phi_j$ may be different from zero for finite $N$. $\mathcal{P}(\lambda^2)$ is a periodic function, but the periodicity will rapidly become unobservable as $N$ increases. As before, we set $\theta_{0,j} = 0$ for all $j$. For simplicity, we also set $|\mathcal{P}_{0,j}| = |\mathcal{P}_0|$ for $j = 1, \ldots, N$. In this case, the model converges to Equation~\ref{burnslab-eq} for large $N$.

In the limit of $\sigma_\phi \lambda^2 << 1$, this system behaves approximately as a Faraday thin source with polarization $|\mathcal{P}_0|$ and $RM$ equal to the mean of the $\phi_j$, which is not exactly zero for finite $N$. This is an analogy of the linear regime of $RM$ jitter, where the $RM$ jitter of a sample of such sources is determined by the scatter in the mean of the $\phi_j$. In the opposite limit, $\sigma_\phi \lambda^2 >>1$, the components add as a more-or-less random set of $N$ vectors on a circle, with expected amplitude of order $|\mathcal{P}_0|/\sqrt{N}$. While there can be specific wavelengths at which the components coincidentally completely depolarize, or completely align, differential Faraday rotation precludes such alignment over a large bandwidth (cf. Figure~\ref{gradient-fig}).  In fact, this system of $N$ Faraday thin components cannot depolarize consistently beyond the statistical level $|\mathcal{P}_0|/\sqrt{N}$. 

\begin{figure*}
\centerline{\resizebox{12cm}{!}{\includegraphics[angle=0,trim={0.0cm 0.5cm 0.0cm 1.5cm}]{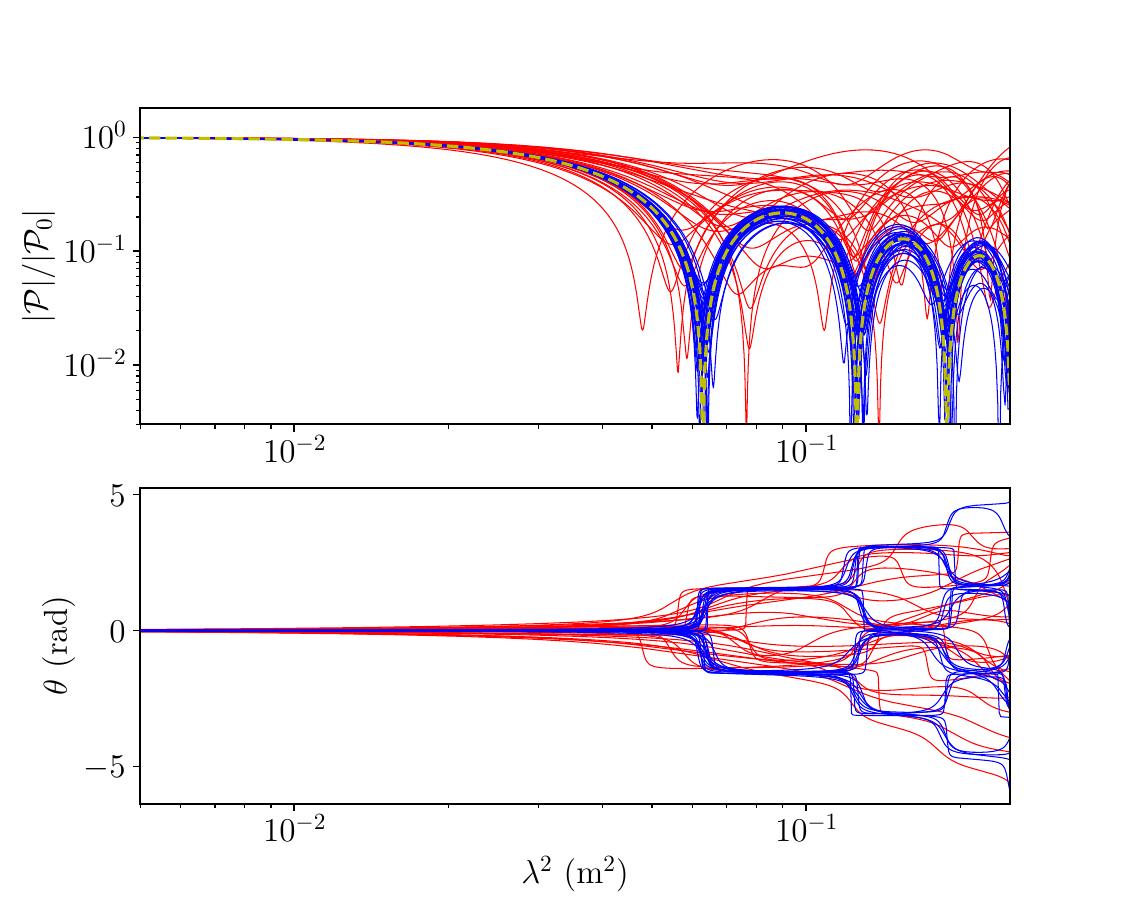}}}
\caption{ Depolarization (top) and $\theta(\lambda^2)$ (bottom) for 30 randomly selected systems with $N = 10$ (red) and $N = 1000$ (blue). The yellow dashed curve in the top panel displays the depolarization of the model in Equation~\ref{burnslab-eq} with $\Phi = 50\ \radm$. The polarization angle of the analytic model (not shown) jumps by $\pi/2$ when it passes through $|\mathcal{P}| = 0$. For this figure, the polarization angle $\theta$ of the models is allowed to vary continuously beyond the range $\left< -\pi/2, \pi/2\right]$. The bifurcations of the curves $\theta(\lambda^2)$ are the effect of incomplete depolarization for finite $N$, which causes the polarization angle to perform an approximate random walk on a grid $k (\pi/2)$, $k= 0, \pm 1, \pm 2, \pm 3, \ldots$ if $N$ is large, but finite.  \label{Ncomp-fig}}
\end{figure*}

Figure~\ref{Ncomp-fig} illustrates the wavelength dependence of the simplified $N$-component model for $N=10$ and $N=1000$, both with a uniform distribution of  $|\phi_j| \le 25\ \radm$. The 10-component models depolarize by a factor $\sim 1/\sqrt{10}$, with a considerable spread, whereas the 1000-component models follow the depolarization of the analytic model ($N \rightarrow \infty$) of Equation~\ref{burnslab-eq} fairly closely.

The dependence of polarization angle on wavelength shows some noteworthy analogies with the models presented in Section~\ref{RMjitter-sec}. The monochromatic rotation measure $\mathcal{R}$ (Equation~\ref{RM-eq}) varies slowly with $\lambda$ until $1/|\mathcal{P}|$ amplification becomes significant. In the $N$-component model, this occurs when relative Faraday rotation of the components has introduced significant spread in their polarization angles within the observed wavelength range.  The analytical limit for $N \rightarrow \infty$ has $\mathcal{R} = 0$, except when $|\mathcal{P}|=0$, where the model passes through the origin in the $QU$ plane and the polarization angle changes by $\pi/2$ \citep{burn1966}.

The $N=1000$ model in Figure~\ref{Ncomp-fig} does not depolarize completely at the nulls of Equation~\ref{burnslab-eq}, but it passes the origin of the $QU$ plane at a short distance, resulting in a large positive or negative $\mathcal{R}$. Subsequent passes by the origin may cause again positive or negative $\mathcal{R}$, giving rise to the braided structure of the blue curves in the bottom panel of Figure~\ref{Ncomp-fig}. The $N=10$ models cannot, in general, depolarize as much. Their wavelength dependence is more erratic, both in depolarization and in $\mathcal{R}$. Systems with smaller $N$ provide an analogy of the larger scale structures in a turbulent plasma within the telescope beam that may cause differential Faraday rotation between a limited number of "cells".

The simulations in Section~\ref{RMjitter-sec} are strictly speaking also of finite $N \sim 5\times 10^5$. Indeed, they do not depolarize below $\sim 1.4 \times 10^{-3} |\mathcal{P}_0|$. This is not a problem because sources with depolarization factor $< 0.01$ are treated as completely depolarized. Simulations at higher resolution confirmed this and gave no indication of different $RM$ jitter within the statistical errors. Surprisingly, some of the models presented in Section~\ref{RMjitter-sec} do show near perfect depolarization at some wavelengths. This is more common when the power spectrum is steep.

We can gain further insight by writing the $RM$ of this system following Equation~\ref{RM-eq}. Writing $\mathcal{P}$ in terms of its real and imaginary components and taking the derivative of $\theta = 0.5 \arctan(U/Q)$ with respect to $\lambda^2$, yields
\begin{widetext} 
\begin{equation}
\mathcal{R} = {1 \over 1 + f}
{\Bigl[\sum \phi_j \cos(2 \phi_j \lambda^2)  \Bigr] \Bigl[ \sum \cos (2 \phi_j \lambda^2)  \Bigr] + \Bigl[\sum \phi_j \sin(2 \phi_j \lambda^2)  \Bigr] \Bigl[ \sum \sin (2 \phi_j \lambda^2)  \Bigr] \over \Bigl[ \sum \cos (2 \phi_j \lambda^2) \Bigr]^2 },
\end{equation}
\end{widetext}
with
\begin{equation}
f = \Bigl[ {\sum \sin (2 \phi_j \lambda^2) \over \sum \cos (2 \phi_j \lambda^2)} \Bigr]^2,
\end{equation}
and all sums are over the index $j = 1 \ldots N$. Consider the limit $\phi_j \lambda^2 << 1$, which is a stronger limit than $\sigma_\phi \lambda^2 << 1$ (as the expectation value of the $\phi_j$ is equal to zero). In this case we can approximate the terms with $\sin$ and $\cos$ in this expression to first order in $\phi_j \lambda^2$ to get the Faraday thin approximation
\begin{equation}
{d \theta \over d \lambda^2} \approx {1\over N} \sum \phi_j  \Bigl[  1 + 4 \sigma_\phi^2 \lambda^4 \Bigr].
\end{equation}
The standard deviation of the mean of $\phi_j$ is $\sigma_\phi/\sqrt{N}$, so we find an expression for the $RM$ jitter
\begin{equation}
\Delta RM = {\sigma_\phi \over \sqrt{N}} \Bigl( 1 + 4 \sigma_\phi^2 \lambda^4 \Bigr).
\end{equation}
In the Faraday thin limit, we find that $RM$ jitter contains a term that is linear in $\sigma_\phi$ and independent of $\lambda$, and a term that is third order in $\sigma_\phi$ that does depend on $\lambda$. Qualitatively similar results were noticed in Figure~\ref{gradient-fig}, and \ref{RMjitter-fig}. The linear term represents the fact that, for finite $N$, the $\phi_j$ average to a non-zero mean that is different for every source in the sample. The third order term represents the effect of differential Faraday rotation between the $N$ components. It is this term that introduces wavelength dependence of $RM$ jitter. $RM$ jitter also reduces as $1/\sqrt{N}$, which means that a plasma that is adequately described by Equation~\ref{depol_model-eq} or Equation~\ref{burnslab-eq} produces no $RM$ jitter.

\section{Discussion}

$RM$ jitter is scatter in $RM$ of a sample of radio sources that arises from variations in Faraday depth related to turbulent structure on the largest angular scales contained within the beam.  Results presented in this paper show that $RM$ jitter is significant compared with measurement errors and the effect of the ionosphere. It constitutes a part of the total extragalactic $RM$ scatter that depends on the wavelength coverage of a survey. For every survey, a linear regime can be identified where $RM$ jitter is the result of unresolved turbulent structure averaging to a non-zero mean Faraday depth, which is independent of wavelength. This is illustrated in Figure~\ref{gradient-fig} by the scatter in the regime $\sigma_\phi \lambda^2 < 1$, and by the different mean Faraday depth of the distributions in Figure~\ref{phi_dist-fig} for simulations with the same power spectrum.

The non-linear regime is characterized by differential Faraday rotation within the beam that gives rise to additional $RM$ scatter. At even higher Faraday depth dispersion, $RM$ jitter enters the saturated regime, where $\Delta RM_{\rm IQR}$ approaches the IQR of the Faraday depth distribution of the source. The Faraday depth dispersion at which the saturated regime is reached, depends on the slope of the power spectrum. Although $RM$ jitter and depolarization are approximately the same for surveys with the same $W_{\rm max}$, the slope of the power spectrum of Faraday depth structure is an additional parameter to be considered.

$RM$ jitter offers an interpretation for the dependence of $RM$ scatter on fractional polarization reported recently by \citet{vanderwoude2024} in terms of a contribution to the total $RM$ scatter that can be attributed to turbulent structure in the near-source plasma. Comparing the scatter in Figure~\ref{fracpol_jitter-fig} with the expected $RM$ errors in POSSUM Low + Mid (Table~\ref{surveys-tab}) suggests that a 3\% threshold on fractional polarization leaves some sources with $RM$ jitter exceeding the measurement errors in the sample. Since this threshold is already near the median fractional polarization of extragalactic radio sources, applying a higher threshold to reduce the scatter would come at the expense of higher rejection rate. The results in Section~\ref{RMjitter-sec} suggest a threshold of 7\% or higher is required to construct a sample in which $RM$ jitter is comparable to $RM$ errors in the surveys considered here. 

The results presented here should be considered preliminary. They show that $RM$ jitter introduces a relation between $RM$ scatter and fractional polarization as observed. More detailed modelling and more data are required for further investigation. Observations of well resolved radio galaxies can provide insight in the power spectrum of Faraday depth structure of the plasma that surrounds radio lobes.
If the link can be confirmed, the sensitivity of $RM$ jitter to the slope of the power spectrum can be exploited to constrain the latter. This would be a new application of the $RM$ grid.

$RM$ jitter has potentially far-reaching implications, because several experiments using the $RM$ grid depend on accurate knowledge of the source-to-source $RM$ scatter. A few examples are discussed briefly in sequence: $RM$ variability, structure functions of $RM$, $RM$ variance from extended extragalactic sources (clusters or galaxies), and $RRM$ as a function of redshift.

$RM$ variability has been observed at low frequencies in compact transient sources, e.g. Fast Radio Bursts \citep[e.g.][]{hilmarsson2021,mckinven2023}, and a Galactic transient \citep{wang2021,weatherhead2024}, and at mm wavelengths in some Active Galactic Nuclei \citep[e.g.][]{goddi2021}. $RM$ variability on time scales of years or decades may exist in some sources. When searching for $RM$ variability using different, historical, surveys, what level of difference between the observed $RM$ can be deemed evidence for variability? This depends on the Faraday depth range in the source and the $\lambda^2$ sampling of the surveys. When comparing $RM$s from different surveys, the correlation of $RM$ jittter between the two surveys (Figure~\ref{RMcorr_2-fig}) must be considered. Older, narrow band, surveys \textit{such as the CGPS and NVSS} are vulnerable to occasional large deviations in $RM$, when $\mathcal{R}$ becomes large in the observed wavelength range. That may render the older data unreliable for variability detection except for very large amplitudes. The effect of $RM$ jitter on \textit{detection} of $RM$ variability can be eliminated with an observing campaign that measures the same wavelength range at all epochs. Also, there should be no concern about sources that are known to have very small $\sigma_\phi$, for example pulsars. Since $RM$ jitter represents scatter in $RM$ from complexity that arises in a turbulent plasma, the \textit{interpretation} of the variability may yet be complicated by this effect.

Structure functions of $RM$ \citep[e.g.][]{simonetti1984,clegg1992,haverkorn2004,stil2011} rely on the difference $RM_i - RM_j$ of different extragalactic sources that are selected to be a certain distance from each other in the sky. The effect of $RM$ jitter on each source pair is different, depending on $\sigma_\phi$ in the sources. This should be uncorrelated with the separation of the sources on the sky, so $RM$ jitter is a power term that is independent of scale, and independent of the common foreground screen, except perhaps in the Galactic plane where a modest Galactic contribution to $\sigma_\phi$ has been detected \citep{livingston2021}. The observed structure function must be corrected for the effect of measurement errors, and an uncorrelated scale-independent extragalactic contribution. $RM$ jitter contributes a survey-specific amount of scatter to the $RM$s of background sources that will affect the slope of the structure function on small scales. The slope of the Galactic $RM$ structure function is shallower than that expected for Kolmogorov turbulence \citep[e.g.][]{haverkorn2004,stil2011}, but this conclusion is sensitive to the corrections discussed here, including $RM$ jitter, which can add non-Gaussian scatter.

Statistical analysis of the observed distribution of $RM$s of distant sources near the line of sight of objects in the foreground has been applied in situations when there is not more than a single background source available for any single target. This work relies on $RM$ scatter because the mean $RM$ of independent objects tends to zero. $RM$ variance has been applied to clusters of galaxies \citep{lawler1982,clarke2001,anderson2021,loi2025}, nearby galaxies \citep{heesen2023,bockmann2023}, the cosmic web \citep{akahori2014}, and quasars with optical absorption lines from intervening systems \citep{bernet2008,bernet2013,farnes2014b}. The statistical significance of any excess scatter in $RM$ depends on the amount of background scatter, and its statistical distribution. Since $RM$ jitter depends on $\lambda^2$ sampling of the survey, and it can be a significant fraction of the total extragalactic $RM$ scatter, it must be evaluated for the survey that is being used in the analysis.

Systematic changes of $RRM$ as a function of redshift may be caused by various interesting cosmological effects, including intervening galactic systems along the line of sight and evolution of the source and its environment over cosmic time. Here too, the information is in scatter of $RRM$ as a function of redshift, $z$ \citep[e.g.][]{hammond2013,aramburo2023}. $RM$ jitter can be a confounding factor, or an object of investigation, in this case because of its predictable dependence on redshift. $RM$ jitter is discussed in this paper in terms of the complete radio source population. Any work that makes a redshift selection should include $(1+z)^{-2}$ for the transformation from observed wavelength to wavelength at the source. The redshift dependence of $RM$ jitter is approximately the redshift dependence of $W_{\rm max}$, which includes $\lambda_{\rm min}$ and $\lambda_{\rm max}$. 

The results presented here reflect also on survey design and programs to establish or monitor sources used for polarization calibration. Faraday rotation of a calibrator measured in one band may only be applicable to another band with an accuracy limited by the expected relative $RM$ jitter. 

The $RM$ jitter phenomenon in itself presents an application of the $RM$ grid that has not previously been considered. As it connects a significant portion of extragalactic $RM$ scatter to Faraday complexity from plasma near the source, it can be applied to investigate the turbulent properties of the near-source medium. While measurements of Faraday complexity are difficult and limited by Faraday depth resolution, measurements of $RM$ are much more robust. Figure~\ref{RMjitter-fig} predicts a certain amount of $RM$ scatter as a function of wavelength-dependent depolarization. The details depend on the slope of the power spectrum, and potentially whether the turbulence is anisotropic. Several authors have reported that polarization of radio sources depends on angular size \citep{cotton2003,grant2010,hales2014,rudnick2014,vernstrom2019,johnston2021}. Observation of a correlation between $RM$ variance and angular size would provide further support for the notion that $RM$ scatter is at least in part related to the circumgalactic medium of the host galaxy.  

The results presented here provide guidance for the conditions when other sources of $RM$ scatter may dominate over $RM$ jitter. Selecting more highly polarized sources reduces the fraction of sources affected by $1/|\mathcal{P}|$ amplification. The $RM$ dispersion found by \citet{vanderwoude2024}  for sources that are less than 3\% polarized would be dominated by $RM$ jitter, whereas the complementary sample displays $RM$ variance that exceeds the predicted $RM$ jitter. The $RM$ variance of sources at low radio frequencies reported by \citet{osullivan2023} is smaller than in POSSUM. The present results indicate this is a selection effect because the LOFAR survey is much more sensitive to depolarization by differential Faraday rotation exceeding $\sim 1\ \radm$, but the resulting scatter still exceeds the expected $RM$ jitter.

\section{Summary and Conclusions}

This paper presents a set of numerical experiments that connect turbulent structure on scales smaller than the beam to variance in $RM$. While this effect was previously noticed by \citet{tribble1991}, this author focused on the median depolarization as a function of wavelength. The results here are more related to the sample variance of radio sources imposed by a turbulent screen with certain power spectrum (amplitude and slope). 

The results of $RM$ synthesis vary significantly between independent realizations of the turbulent screen {\it with the same statistical parameters}, affecting both the $RM$ and metrics of Faraday complexity. This effect is referred to as $RM$ jitter. The range of possible $RM$ exceeds the true range of Faraday depths in the screen, even though the screen represents a continuous distribution of Faraday depths. $\Delta RM_{\rm IQR}$ of a plasma with a turbulent power spectrum with slope $\gamma = -2.5$ varies between $\Delta RM_{\rm IQR} \sim \sigma_\phi/5$ when depolarization by differential Faraday rotation is insignificant (the linear regime), and the inter-quartile radio of a Gaussian with standard deviation $\sigma_\phi$ when depolarization by differential Faraday rotation is very significant (the saturated regime). In between these two limits, $RM$ jitter depends on the wavelength coverage of a survey (the non-linear regime).

$RM$ jitter is independent of Faraday depth resolution. The non-linear increase of $RM$ jitter with $\sigma_\phi$ is a depolarization effect, where $1/|\mathcal{P}|$ amplification of Faraday rotation in the $QU$ plane occurs. This is a re-statement of what \citet{tribble1991} considered the $\sigma_\phi \lambda^2 = 1$ boundary in the context of quasi-monochromatic data. For most modern, broad band, surveys, the depolarization parameter $W_{\rm max}$ introduced by \citet{rudnick2023} is a helpful parameter to assess the amount of $RM$ jitter in a survey.

When comparing $RM$s derived from different surveys, $RM$ jitter can introduce scatter that exceeds measurement errors, in particular if at least one survey used a small bandwidth, such as the NVSS or CGPS. $RM$s from different surveys are highly correlated in the linear regime of $RM$ jitter, loosely correlated with a slope different from 1 in the non-linear regime, and virtually uncorrelated in the saturated regime of $RM$ jitter.

Results in this paper show that it is useful to distinguish between the concepts $RM$, derived from $RM$ Synthesis or $QU$ fitting, Faraday depth $\phi$, which is proportional to the integral of electron density and line-of-sight component of the magnetic field along the line of sight, and $\mathcal{R}$, the monochromatic slope of $\theta(\lambda^2)$. These quantities can all be very different, depending on circumstances.

$RM$ jitter affects several applications of the $RM$ grid. It must be taken into account when $RM$ differences are considered. This includes but is not restricted to investigations of $RM$ variability (comparing results of $RM$ surveys), $RM$ structure functions, statistical analysis of $RM$ variance as a function of impact parameter, and $RRM$ as a function of redshift. Science that depends on $RM$ scatter over a wide area of sky, would benefit from uniform $\lambda^2$ coverage. Searches for $RM$ variability benefit significantly from consistent $\lambda^2$ coverage.

\begin{acknowledgements}
The author thanks L. Rudnick for helpful comments on the manuscript prior to submission and the anonymous referee for constructive in-depth feedback. The author acknowledges the support of the Natural Sciences and Engineering Research Council of Canada (NSERC), 2019-04848, and a generous private donor of a substantial server for data processing and simulations. The author thanks Solveig Thompson for her helpful solutions to python coding for the on-line material.
\end{acknowledgements}

\appendix

This appendix presents some animated figures that illustrate simulated sources as a function of wavelength. One of the most striking features of these models is that, as the power spectrum becomes steeper, from $\gamma=-2.0$ to $\gamma=-3.0$, the tracks that models with $\phi_0 = 0\ \radm$ follow in the $QU$ plane evolve from mostly radial to strongly curved. Models with uniform $\theta_0$ show structure in $Q$ and $U$ that mimics the Faraday depth structure of the screen when $\sigma_\phi \lambda^2 << 1$.  When $\sigma_\phi \lambda^2 \gtrsim 0.5$, the structure in $Q$ and $U$ changes as regions with the strongest Faraday rotation display substantial changes in polarization angle. As the structure in $Q$ and $U$ breaks up into smaller positive and negative regions, the integral over the source converges to zero and significant depolarization sets in. Azimuthal excursions of the model track closer to the origin result in larger $|\mathcal{R}|$, and significant changes in $\mathcal{R}$ with wavelength. This source has a maximum $\mathcal{R} = 24.9\ \radm$ at 1.2 GHz. The reader is encouraged to explore the variation of $\mathcal{R}$ across L band (1-2 GHz) or S band (2 - 4 GHz) sliding through the animation with a mp4 viewer.

\begin{figure*}
\centerline{\resizebox{\textwidth}{!}{\includegraphics[angle=0,trim={0.0cm 0.0cm 0.0cm 0.5cm}]{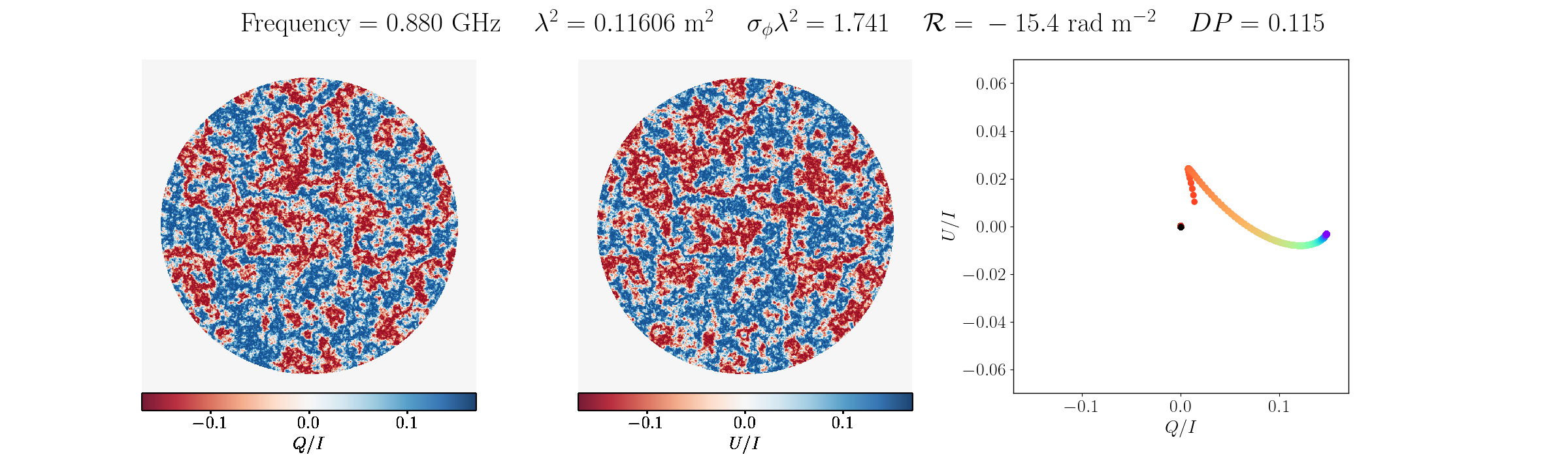}}}
\caption{ Example of a simulated source with uniform intrinsic polarization angle $\theta_0 = 0\ \radm$, $\sigma_\phi = 15\ \radm$, $\gamma=-3.0$ and uniform $|\mathcal{P}_0|$ = 0.15, $\phi_0 = 0.0\ \radm$. The left panel shows $q = Q/I$ across the source.  The middle panel shows $u = U/I$ across the source. The right panel shows integrated $Q/I$ and $U/I$ starting at frequency 4 GHz (blue) down to the frequency shown at the top (880 MHz, red). The complete animated sequence in the on-line figure spans the frequency range from 4 GHz to 600 MHz. The black dot marks the origin, $|\mathcal{P}| = 0$. The depolarization factor $DP$ is defined as $|\mathcal{P}|/|\mathcal{P}_0|$. $\mathcal{R}$ changes sign a few times before this source depolarizes completely. Note the difference in scale between the $Q/I$ and $U/I$ axes of the rightmost panel, which affects the displayed distance of the model from the origin. The spectral resolution of the simulations analyzed in this paper is ten times better than the spectral resolution shown here, and the highest frequency of the simulations is $8\ \rm GHz$. An animated figure that covers the frequency range 4 GHz to 600 MHz is available in the ancillary files.  \label{anim_0302-fig}}
\end{figure*}

\begin{figure*}
\centerline{\resizebox{\textwidth}{!}{\includegraphics[angle=0,trim={0.0cm 0.0cm 0.0cm 0.5cm}]{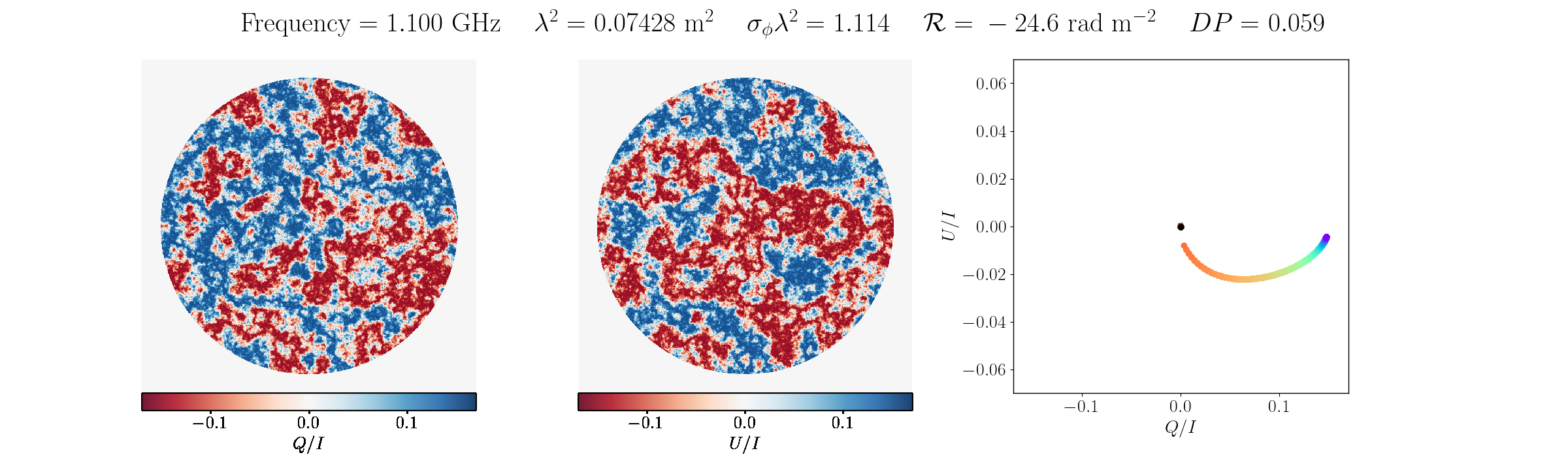}}}
\caption{ Example of a simulated source with uniform intrinsic polarization angle $\theta_0 = 0\ \radm$, $\sigma_\phi = 15\ \radm$, $\gamma=-3.0$ and uniform $|\mathcal{P}_0|$ = 0.15, $\phi_0 = 0.0\ \radm$.  The Faraday screen of this source has the same statistical parameters as the screen in Figure~\ref{anim_0302-fig}, but $\mathcal{R}$ is predominantly negative here. Note that at 1100 MHz, this source is more depolarized than the source in Figure~\ref{anim_0302-fig}, even through $\sigma_\phi \lambda^2$ is smaller here. An animated figure that covers the frequency range 4 GHz to 600 MHz is available in the ancillary files.   \label{anim_0385-fig}}
\end{figure*}

\begin{figure*}
\centerline{\resizebox{\textwidth}{!}{\includegraphics[angle=0,trim={0.0cm 0.0cm 0.0cm 0.5cm}]{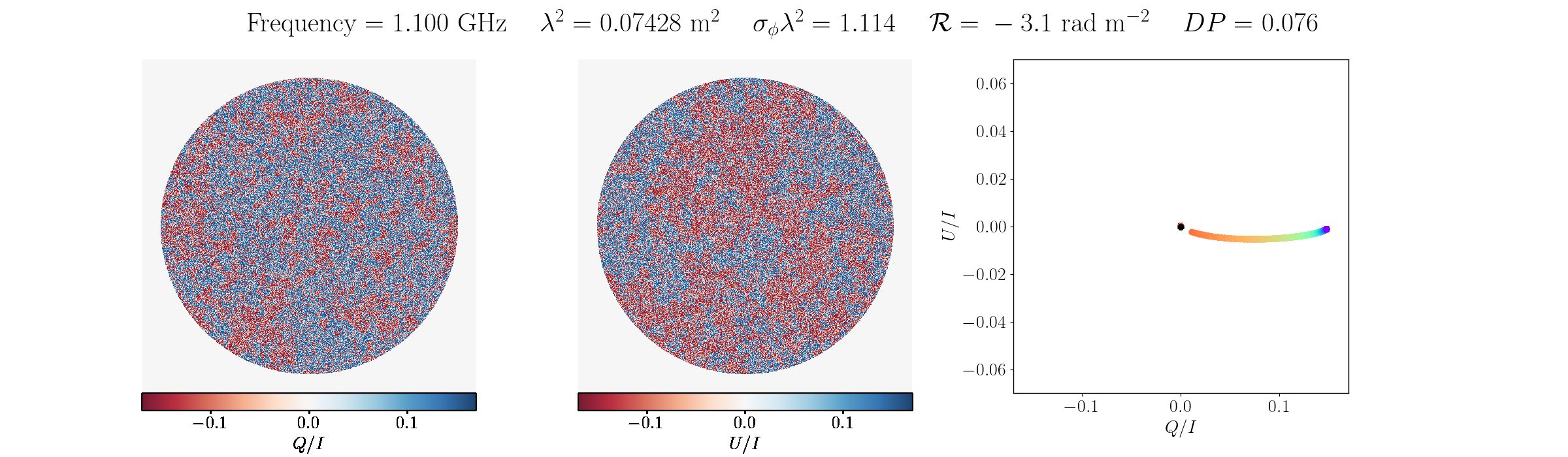}}}
\caption{ Example of a simulated source with uniform intrinsic polarization angle $\theta_0 = 0\ \radm$, $\sigma_\phi = 15\ \radm$, $\gamma=-2.0$ and uniform $|\mathcal{P}_0|$ = 0.15, $\phi_0 = 0.0\ \radm$.  Compared with Figure~\ref{anim_0302-fig} and Figure~\ref{anim_0385-fig}, there is much less power on large scales, and the path of this source toward the origin is much more radial. As a result, $\mathcal{R}$ is much closer to zero until depolarization is almost complete (see also Figure~\ref{fracpol_jitter-fig} and Figure~\ref{Ncomp-fig}). This is characteristic of models with $\gamma=-2.0$. An animated figure that covers the frequency range 4 GHz to 600 MHz is available in the ancillary files.  \label{anim_0333-fig}}
\end{figure*}

In this paper, the intrinsic polarization angle of the synchrotron emission is assumed uniform across the source, $\theta_0 = 0$.  Modelling non-uniform $\theta_0$ requires further postulates about possible values and spatial structure of the intrinsic polarization angle. To gain insight in the effect of non-uniform $\theta_0$ on the results presented here, consider first an infinitesimal section of the source, or, in the discrete case of Section~\ref{Ncomp-sec}, a single Faraday thin component. For simplicity, we write both cases as a countable set with index $j$. Split the complex polarization of this part of the source as the \textit{sum} of the complex polarization at $\lambda=0$ and a wavelength-dependent term,
\begin{equation}
\mathcal{P}_j = \mathcal{P}_{0,j} + \mathcal{P}_{\lambda,j}.
\label{onecomp-eq}
\end{equation}
This sum appears contradictory to more common expressions like Equation~\ref{depol_model-eq} and Equation~\ref{burnslab-eq}, but mathematically, this separation into two terms can always be defined. It is equivalent with writing $\mathcal{P}_j$ as a vector in the $Q$,$U$ plane and then writing this vector as the sum of a constant vector that represents the intrinsic polarization at $\lambda = 0$ and a vector that depends on wavelength. By definition, $\mathcal{P}_{\lambda,j}=0$ for $\lambda=0$. Integration over the source, or summation over $N$ Faraday components, yields the total wavelength-dependent complex polarization of the source, as the sum of the integrals of the two terms in Equation~\ref{onecomp-eq},
\begin{equation}
\mathcal{P}_{\rm tot} = \mathcal{P}_{0,\rm tot} + \mathcal{P}_{\lambda, \rm tot}.
\label{sumcomp-eq}
\end{equation}
All of the intrinsic polarization of the source, including the intrinsic angle distribution, is included in $\mathcal{P}_{0, \rm tot}$. All of the Faraday rotation is included in $\mathcal{P}_{\lambda, \rm tot}$. This illustrates that the intrinsic polarization angle distribution defines the starting point of the model in the $Q$,$U$ plane. Because the two terms in Equation~\ref{sumcomp-eq} add up as 2-dimensional vectors \citep[c.f.][their Figure 1]{brentjens2005}, and one is subject to Faraday rotation, a different starting point can change the shape of the track of a source in the $Q$,$U$ plane.

\begin{figure*}
\centerline{\resizebox{\textwidth}{!}{\includegraphics[angle=0,trim={0.0cm 0.0cm 0.0cm 0.5cm}]{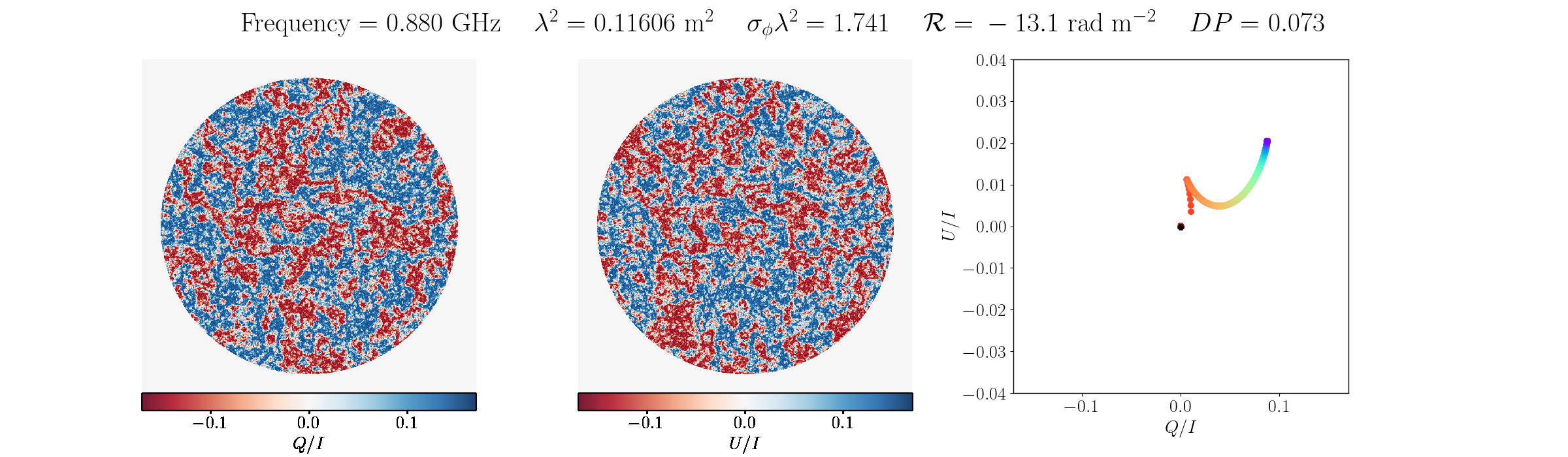}}}
\caption{ Repeat of the simulated source in Figure~\ref{anim_0302-fig} with the same Faraday screen, including an intrinsic polarization angle distribution as described in the text.  The depolarization parameter $DP$ now includes the wavelength-independent depolarization effect of the intrinsic polarization angle distribution. Note the scale of the right panel is different from Figure~\ref{anim_0302-fig}. An animated figure that covers the frequency range 4 GHz to 600 MHz is available in the ancillary files. \label{anim_0302_angle-fig}}
\end{figure*}

Figure~\ref{anim_0302_angle-fig} illustrates the effect of intrinsic polarization angle structure on the simulated source shown in Figure~\ref{anim_0302-fig}. The adopted intrinsic polarization angle distribution is defined by a random field with a power law index $-2.5$ and cut-off at the scale of the source. The standard deviation of $\theta_0$ across the source is $0.5\ \rm rad$, and the expectation value of $\theta_0$ (not the mean of the simulated angles) is zero.  The polarization angle structure is independent of the structure of the Faraday screen. Since $dQ/d\theta = 0$ when $\theta=0$, most of the intrinsic angle structure at short wavelengths appears in the Stokes $U$ image. Integration over the source yields intrinsic polarization with $|U| << |Q|$. The models with uniform $\theta_0 = 0$ have $U=0$ at $\lambda=0$.

The $Q$ and $U$ images in the animated figure illustrate the principle outlined in Equation~\ref{sumcomp-eq}.  Comparing the track in the $Q$,$U$ plane in Figure~\ref{anim_0302_angle-fig} with that of the sources with uniform $\theta_0$ in Figure~\ref{anim_0302-fig} and Figure~\ref{anim_0385-fig}, shows that the effect of an independent realization of the Faraday screen with the same power spectrum is much larger than changing the intrinsic polarization angle distribution. Several more experiments support this conclusion. The adopted polarization angle distribution does distort the track, affecting $\mathcal{R}$, but it is not expected to fundamentally change the statistics of $RM$ of a sample of independent sources beyond the $1/|\mathcal{P}|$ depolarization effects discussed in this paper.

In Figure~\ref{anim_0302_angle-fig}, the structure in $Q$ and $U$ at short wavelengths is defined by the intrinsic polarization angle structure. The structure in the Stokes $U$ image  in Figure~\ref{anim_0302-fig} at short wavelengths mimics the structure of the Faraday screen because of a modest amount of Faraday rotation. Comparing the animated Figure~\ref{anim_0302-fig} with Figure~\ref{anim_0302_angle-fig} at longer wavelengths shows similar wavelength-dependent structures in the $Q$ and $U$ images associated with locations of more extreme Faraday depth.

\begin{figure*}
\centerline{\resizebox{12cm}{!}{\includegraphics[angle=0,trim={0.0cm 0.2cm 0.0cm 1.5cm}]{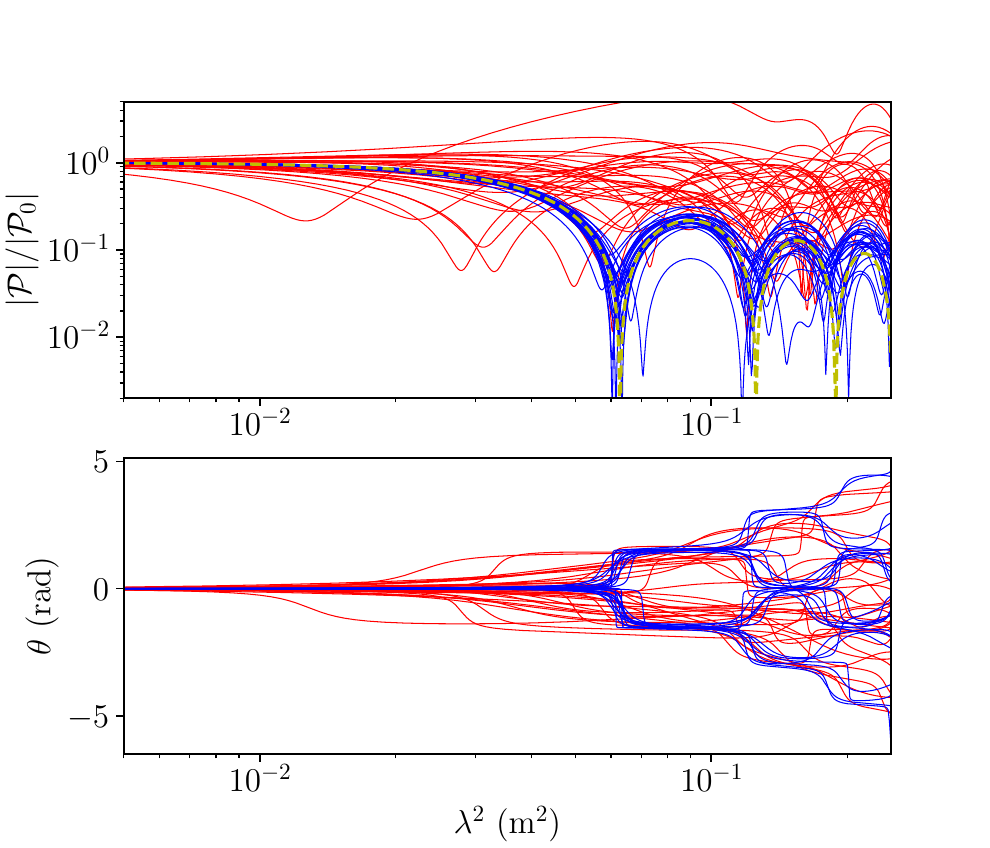}}}
\caption{ Repeat of the experiment in Figure~\ref{Ncomp-fig} showing 30 independent sources with $N=10$ (red) and 30 sources with $N=1000$ (blue) Faraday thin components, now with $\theta_{0,j}$ ($j = 1, \ldots, N$) drawn from a uniform distribution between $-1$ and $1$ rad. The dashed yellow curve is the same as in Figure~\ref{Ncomp-fig}.  \label{Ncomp_rand-fig}}
\end{figure*}

Further insight can be gained from the models in Section~\ref{Ncomp-sec}. Figure~\ref{Ncomp_rand-fig} repeats Figure~\ref{Ncomp-fig} for models where the intrinsic polarization angle $\theta_{0,j}$ of each of the $N$ components was drawn at random from a uniform distribution between $-1$ and $1$ rad. The models were normalized relative to their fractional polarization in the high-frequency limit, and the actual mean of the $\theta_{0,j}$ was subtracted from the angle of every source to accommodate comparison of independent sources. Note how the main features of Figure~\ref{Ncomp-fig} are preserved, although the $N=10$ curves appear to diverge more at shorter wavelengths.  For small $N$, coincidental depolarization may require less differential Faraday rotation if the individual Faraday components are already spread out in the $Q$,$U$ plane at shorter wavelengths. There is more significant repolarization of sources in Figure~\ref{Ncomp_rand-fig} \citep[c.f.][]{farnes2014a}. There may be analogies of this behaviour in more complex systems. Detailed modelling of the intrinsic polarization angle distribution is beyond the scope of this paper, but even the suggestion that it may be a complication in our understanding of $RM$ scatter in surveys is noteworthy.

{}

\end{document}